\documentclass[aps,pra,twocolumn,superscriptaddress,hyperref,showpacs]{revtex4-1}
\usepackage{graphicx}
\usepackage{amsfonts}
\usepackage{natbib}
\usepackage{graphicx}
\usepackage{color}
\usepackage{changes}

\begin{document}


\title{Universal Steps in Quantum Dynamics with Time-Dependent Potential-Energy Surfaces: Beyond the Born-Oppenheimer Picture}


\author{Guillermo Albareda}
\email[]{albareda@ub.edu}
\affiliation{Departament de Qu\'imica F\'isica and Institut de Qu\'imica Te\`orica i Computacional, Universitat de Barcelona, 08028 Barcelona, Spain.}

\author{Ali Abedi}
\email[]{aliabedik@gmail.com}
\affiliation{Nano-Bio Spectroscopy Group and ETSF, Universidad del Pa\'is Vasco, UPV/EHU, 20018 San Sebastián, Spain.}
\author{Ivano Tavernelli}
\email[]{ita@zurich.ibm.com}
\affiliation{IBM Research GmbH, Z\"urich Research Laboratory, 8803 R\"uschlikon, Switzerland.}
\author{Angel Rubio}
\email[]{angel.rubio@mpsd.mpg.de}
\affiliation{Nano-Bio Spectroscopy Group and ETSF, Universidad del País Vasco, CFM CSIC-UPV/EHU , 20018 San Sebastián, Spain.}
\affiliation{Max Planck Institute for the Structure and Dynamics of Matter, Luruper Chaussee 149, 22761 Hamburg, Germany.}

date{\today}

\begin{abstract}
It was recently shown [G. Albareda, et al., Phys. Rev. Lett. 113, 083003 (2014)] that within the conditional decomposition approach to the coupled electron-nuclear dynamics, 
the electron-nuclear wave function can be exactly decomposed into an ensemble of nuclear wavepackets effectively governed by nuclear conditional time-dependent potential-energy surfaces ($\mathbb{C}$-TDPESs). 
Employing a one-dimensional model system we show that for strong nonadiabatic couplings the nuclear $\mathbb{C}$-TDPESs exhibit steps that bridge piecewise adiabatic Born-Oppenheimer PESs.
The nature of these steps is identified as an effect of electron-nuclear correlation. Furthermore,  
a direct comparison with similar discontinuities recently reported in the context of the exact factorization framework allows us to draw conclusions about the universality of these discontinuities, viz.
they are inherent to all nonadiabatic nuclear dynamics approaches based on (exact) time-dependent potential energy surfaces.
\end{abstract}


\maketitle

\section{Introduction}
The description of the correlated electron-nuclear dynamics remains a formidable challenge in condensed-matter physics and theoretical chemistry~\cite{exp1,exp2,exp3,exp4,exp5,exp6}. 
Relying on the Born-Huang expansion of the full molecular wave function, a majority of approaches that provide a numerically accurate description of the so-called ``nonadiabatic'' processes require the propagation of a set 
of many-body nuclear wavepackets on a coupled set of Born-Oppenheimer potential-energy surfaces (BOPESs)~\cite{rev1,rev2,rev3,rev4,rev5,truhlar,Tavernelli2,Tavernelli3}. 
Whenever electron-nuclear coherence effects are unimportant, efficient mixed quantum-classical propagation techniques can be used~\cite{SH,tully1998mixed,kapral1999mixed,persico2014overview,SawadaMetiu,Martinez,truhlar_ants}.
However, in order to account for strong correlations, the access to quantum features of the nuclear motion such as wavepacket spreading, interferences, tunneling or splitting is crucial. 
Hence, a reliable description of molecular dynamics becomes very expensive due to the calculation of the full set of BOPESs, a (time-independent) problem that scales 
exponentially with both the nuclear and electronic degrees of freedom~\cite{BurghardtCederbaum,MCTDH1,MCTDH2}. 

\begin{figure*}
\includegraphics[width=\linewidth]{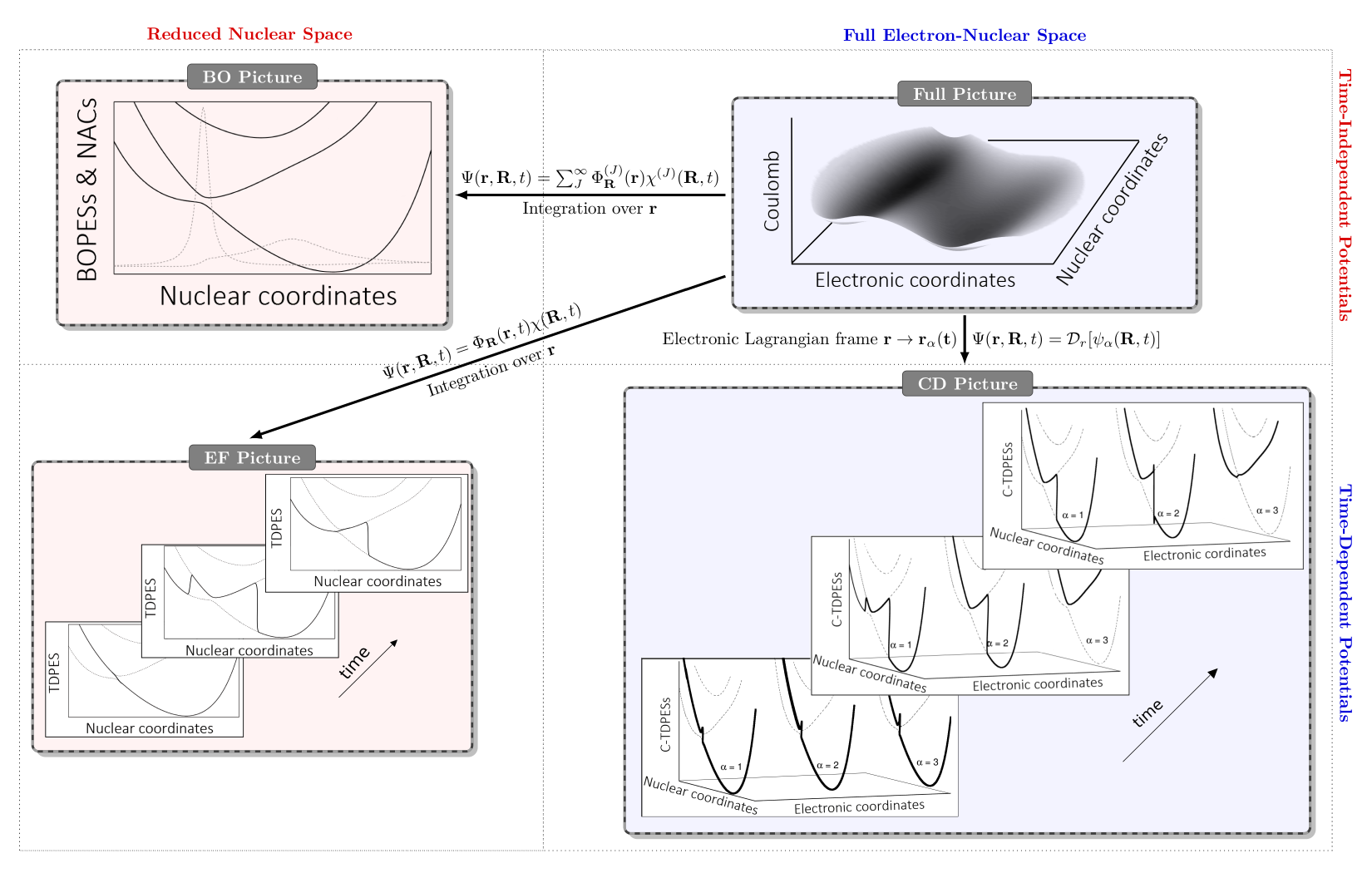}
\caption{Schematic representation of the four (exact) approaches to correlated electron-nuclear dynamics that will be discussed in this work. 
We will be considering the following representations of the molecular wave function: (upper-right panel) full electron-nuclear wave function, 
(upper-left panel) Born-Huang expansion of the electron-nuclear wave function, (bottom-left) exact factorization of the electron-nuclear wave function, and (bottom-right) conditional decomposition of the 
electron-nuclear wave function. These four pictures are respectively referred to as \textit{Full Picture}, \textit{BO Picture}, \textit{EF Picture}, and \textit{CD Picture}.
The nuclear equations of motion that arise in each representation are mathematically very different.
In particular, we distinguish between approaches where the nuclear dynamics occurs in a reduced nuclear subspace (viz. BO and EF Pictures) and approaches where it occurs in the full electron-nuclear Hilbert space (viz. Full and CD Pictures).
Further, we distinguish between approaches where the (effective) nuclear potential-energy surfaces are time-independent (viz. Full and BO Pictures) and approaches where these are time-dependent (viz. EF and CD Pictures).
For the sake of simplicity, we omitted any external field in the scheme.}
\label{fig_schematic}
\end{figure*}

Towards an alternative description of the coupled electron-nuclear dynamics beyond the Born-Oppenheimer (BO) picture (i.e. the Born-Huang expansion of the molecular wave function),  
two different formally exact frameworks, viz. the exact factorization (EF)~\cite{Ali1} and the conditional decomposition (CD)~\cite{Guil1}, have been recently proposed.
These two alternative approaches are to be added to the already existing frameworks that avoid the BO picture of nonadiabatic dynamics~\cite{alonso2008Ehrenfest,mceniry2010Ehrenfest,horsfield2004Ehrenfest}.
In the EF and the CD approaches the nuclear dynamics takes place on time-dependent potential energy surfaces.
This is in contrast with the BO picture where the nuclear wavepacket, with contributions on several BOPESs, undergoes transitions in the regions of strong nonadiabatic coupling (NAC) (see Fig.~\ref{fig_schematic}).
The EF and the CD approaches are, however, grounded on very different mathematical frameworks.
In the EF approach, the electron-nuclear wave function is written as a direct product of electronic and nuclear wave functions. 
The resulting nuclear equations of motion arise from Frenkel's stationary action principle and do involve the integration of the electronic degrees of freedom~\cite{Ali1}.
Alternatively, in the CD approach electrons and nuclei are treated on an equal mathematical footing and hence their description is kept at the full configuration level (see Fig.~\ref{fig_schematic}).
After a partial time-dependent coordinate transformation is applied to the time-dependent Schr\"odinger equation (TDSE), the nuclear (electronic) coordinates are propagated
along with the electronic (nuclear) wave function such that the new coordinates remain located where the full molecular wave function has a significant amplitude~\cite{Guil1}.

A major distinctive characteristic of the CD approach is the fact that it avoids the integration of the electronic degrees of freedom. 
It constitutes in this respect a safe starting point to draw new connections between other formally exact frameworks.
In this work we aim at investigating the generic features of the exact conditional (nuclear) time-dependent potential-energy surfaces ($\mathbb{C}$-TDPESs) that arise in the CD approach in the presence of strong non-adiabatic couplings.
This is a necessary study towards the use of the CD method in real scenarios and will serve as the basis for future development of practical approximations to the nuclear $\mathbb{C}$-TDPESs.
A major result is that, besides the radically different natures of the CD and EF frameworks, the exact $\mathbb{C}$-TDPESs exhibit discontinuous steps connecting different BOPESs analogous to a paradigmatic feature 
of the effective time-dependent potential that governs the nuclear dynamics in the EF framework~\cite{Ali1}.
By establishing a formal connection between the CD and the EF frameworks we will elaborate on the universality of these discontinuities. 
We will be showing that the paradigm shift associated with the transition from the many BOPESs and NACs 
to the (single) time-dependent potential energy surface entails a discontinuous behavior of the resulting effective potentials 
during the branching of the nuclear probability density in nonadiabatic electronic transitions. 

This paper is organized as follows. In Section \ref{theory}, we introduce the CD approach to electron-nuclear correlated dynamics and describe its main mathematical ingredients. 
In Section \ref{BOPES} we will provide some physical intuition on the role played by the $\mathbb{C}$-TDPESs in connection with the BOPESs of the BO picture.
Numerical simulations are thereafter, in Section \ref{numerical}, used to investigate the main features of the exact $\mathbb{C}$-TDPESs in the presence of strong nonadiabatic couplings. 
In view of the results, in Section \ref{discussion} we will establish a formal connection between the CD and the EF approaches and elaborate on the universality of the discontinuous behavior of the resulting effective potentials 
in fully time-dependent approaches to electron-nuclear coupled dynamics. 
Conclusions will be drawn in Section \ref{conclusion}.

\section{The Conditional Decomposition}\label{theory}
\begin{figure*}
\includegraphics[width=0.9\textwidth, height=8cm]{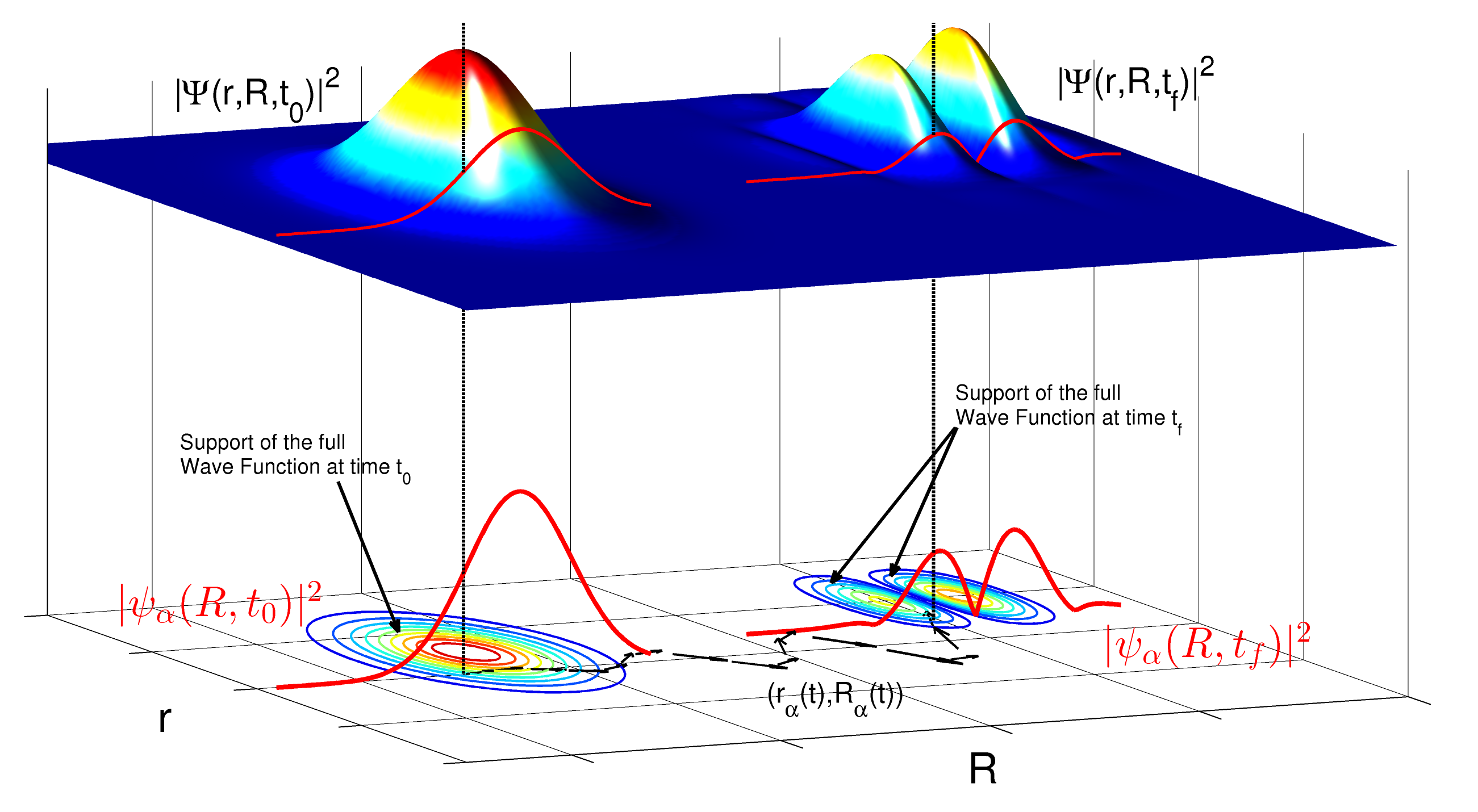}
\caption{Schematic representation of the Conditional Decomposition approach to molecular dynamics for a 2-dimensional system. The full nuclear probability-density $|\Psi(r,R,t)|^2$ 
is plotted at two different times $t_0$ and $t_f$, together with a conditional amplitude $|\psi_\alpha(R,t)|^2$ (in red) for a
particular trajectory $\{r_\alpha (t), R_\alpha (t)\}$. Black arrows denote the velocity field $\{\dot{r}_\alpha (t), \dot{R}_\alpha (t)\}$, and contour plots of the full nuclear wave function are also
shown for clarity.}
\label{fig_0}
\end{figure*}

Throughout this manuscript we use atomic units, and electronic and nuclear coordinates are collectively denoted by $\mathbf{r} = \{ \mathbf{r}_1,..,\mathbf{r}_{N_e} \}$ and 
$\mathbf{R} = \{ \mathbf{R}_1,..,\mathbf{R}_{N_n} \}$, where $N_e$ and $N_n$ are the total number of electrons and nuclei.
We have recently proved~\cite{Guil1} that the full (non-relativistic) electron-nuclear wave function $\Psi(\mathbf{r},\mathbf{R},t)$ can be exactly decomposed into an ensemble of conditional nuclear wave functions defined as:
\begin{equation}
 \psi_\alpha(\mathbf{R},t) := \Psi(\mathbf{r}_\alpha(t),\mathbf{R},t).
\end{equation}
Provided that the electronic trajectories 
$\{ \mathbf{r}_\alpha(t) \equiv \mathbf{r}_{1,\alpha}(t),..,\mathbf{r}_{N_e,\alpha}(t)\}$ explore the electronic support of $|\Psi|^2$
at any time~\footnote{The electronic support of the full probability density $|\Psi(\mathbf{r},\mathbf{R},t)|^2$ can be defined as:
$\textit{supp}(|\Psi(\mathbf{r},\mathbf{R},t)|^2) = \{ \mathfrak{r} \in \mathbf{r} \; \mathbf{|} \; |\Psi(\mathbf{r},\mathbf{R},t)|^2 > \lambda  \}$, where $\lambda$ denotes a given lower bound of the probability density.}, 
these conditional wave functions can be always used to reconstruct the full wave function~\footnote{{
This can be easily done from $\Psi(\mathbf{r}, \mathbf{R} ,t) = {\mathcal{D}}_\mathbf{r} [\psi_\alpha]$, where the transformation ${\mathcal{D}}$ has been defined in Eq.~(\ref{transform}).}} 
(or equivalently any observable).
Notice, however, that since the initial conditions of a trajectory-based simulation can be generated 
with importance-sampling techniques, conditional decompositions can be exploited to circumvent the problem of storing and propagating a many-particle wave function~\cite{Guil1,JPCL,PRBGuille,JCE,PRLOriols,albareda2016bitlles}. 
The conditional wave function concept is illustrated in Fig.~\ref{fig_0} for a simple two-dimensional (one electron plus one nucleus) system.

Throughout this work we will use quantum trajectories $\mathbf{r}_{\xi,\alpha}(t)$
and $\mathbf{R}_{\nu,\alpha}(t)$, that are defined through ~\cite{AppBohm}: 
\begin{equation}
\dot{\mathbf{r}}_{\xi,\alpha}(t) = \mathbf{\nabla}_\xi S|_{\{\mathbf{r}_\alpha(t),\mathbf{R}_\alpha(t)\}}, 
\end{equation}
and
\begin{equation}
  \dot{\mathbf{R}}_{\nu,\alpha}(t) = {\mathbf{\nabla}_\nu S}/{M_\nu} |_{\{\mathbf{r}_\alpha(t),\mathbf{R}_\alpha(t)\}},
\end{equation}
where $S$ is the phase of the full electron-nuclear wave function $\Psi = |\Psi|e^{iS}$.
Note that the choice of quantum (hydrodynamic) trajectories is not mandatory. Alternatively, trajectory-based Monte Carlo or importance-sampling techniques can be used, provided that they sample the (time-dependent) 
quantum-probability density.
It is however useful for the purpose of this work to choose Bohmian trajectories because they do sample the support of the full wave function by definition~\cite{BookOriols,AppBohm}.
For the sake of simplicity, hereafter we will omit the explicit dependence of the trajectories on time, i.e. $\left\{\mathbf{r}_{\alpha}\right\} \equiv \left\{\mathbf{r}_{\alpha}(t)\right\}$. 

In the absence of time-dependent external vector fields, the system is described by the Hamiltonian:
\begin{equation}
 \hat H = \hat T_e(\mathbf{r}) + \hat T_n(\mathbf{R}) +  W(\mathbf{r},\mathbf{R},t),
\end{equation} 
where $\hat T_e=-\sum\nolimits_{\xi=1}^{N_e} {{\nabla}^2_\xi}/{2}$ and $\hat T_n=-\sum\nolimits_{\nu=1}^{N_n} {\nabla^2_\nu}/{2M_\nu}$ 
are the electronic and nuclear kinetic energy operators, and 
$W(\mathbf{r},\mathbf{R},t) = W_{ee}(\mathbf{r}) + W_{nn}(\mathbf{R}) + W_{en}(\mathbf{r},\mathbf{R})$ denotes the Coulombic interactions. 
The wave functions  $\psi_\alpha(\mathbf{R},t)$ then obey the following equations of motion~\cite{Guil1}:
\begin{equation}\label{cond_1}
  i d_t\psi_\alpha(\mathbf{R},t)  = \Big\{  \hat T_{n}(\mathbf{R})  +  V_\alpha(\mathbf{R},t)  \Big\} \psi_\alpha(\mathbf{R},t), 
\end{equation}
where the effective potentials,
\begin{equation}
\label{Valpha}
 V_\alpha(\mathbf{R},t)= W_\alpha(\mathbf{R},t) + K_\alpha(\mathbf{R},t) + A_\alpha(\mathbf{R},t),
\end{equation}
are named conditional time-dependent potential-energy surfaces ($\mathbb{C}$-TDPESs). 
Each $\mathbb{C}$-TDPES consists of three terms: the Coulombic potential,
\begin{equation}\label{bear_coulomb}
  W_\alpha(\mathbf{R},t) = W(\mathbf{r},\mathbf{R},t)|_{\mathbf{r}_\alpha},
\end{equation}
the kinetic correlation potential: 
\begin{equation}\label{kinetic_corr}
 K_\alpha(\mathbf{R},t) = \frac{\hat T_e \Psi}{\Psi} \Big|_{\mathbf{r}_\alpha}, 
\end{equation}
and a pure advective potential, 
\begin{equation}\label{advective_corr}
 A_\alpha(\mathbf{R},t) = i\sum_{\xi=1}^{N_e} \frac{\mathbf{\nabla}_\xi \Psi}{\Psi} \Big|_{\mathbf{r}_\alpha} \cdot \dot{ \mathbf{r}}_{\xi,\alpha}.
\end{equation}
Here we emphasis that unlike the parametrized nuclear static potentials that arise in the BO picture or the effective exact time-dependent potential in the EF approach to electron-nuclear dynamics,
the $\mathbb{C}$-TDPESs of the CD method in Eqs.~(\ref{bear_coulomb})-(\ref{advective_corr}) do not involve the tracing-out of the electronic degrees of freedom.
The $\mathbb{C}$-TDPESs do live in the full electron-nuclear configuration space, and therefore they cannot be defined (strictly speaking) as potential-energy surfaces.
The $\mathbb{C}$-TDPESs consists of three terms, viz. the bear Coulomb potential plus two additional fictitious potentials, i.e. the kinetic and the advective correlations. 
The advective correlation potential represents the rate of change of the nuclear conditional wave functions due to the time-dependence of the new electronic coordinates (the Lagrangian trajectories). 
Instead, the kinetic correlation potential defines a viscous/drag-like quantum effect of the electrons on the nuclear degrees of freedom.

%

As shown in Eq.~(\ref{cond_1}), the evolution of each nuclear wave function $\psi_\alpha(\mathbf{R},t)$ is, in general, non-unitary due to the complex nature of the kinetic and advective potentials. 
Therefore, the conditional wave functions  $\psi_\alpha(\mathbf{R},t)$ do not obey a normalization condition individually, but rather a collective normalization constraint given by:
\begin{equation}\label{overall}
 \int d\mathbf{R}\int|\Psi(\mathbf{r},\mathbf{R},t)|^2 d\mathbf{r} = \int d\mathbf{R}\int \mathcal{D}_\mathbf{r}(|\psi_\alpha|^2) d\mathbf{r} = 1,
\end{equation}
where the transformation ${\mathcal{D}}$ has been defined as~\footnote{Notice that the summation over $\alpha$ implies the discretization of the electronic subspace into 
a countable (infinite) number of points. While the support of $ \Psi(\mathbf{r},\mathbf{R},t)$ consists of an countless (infinite) number of points, we have assumed here the discretization of the configuration space on a numerical grid.}:
{\begin{equation}\label{transform}
 {{\mathcal{D}}_\mathbf{a} [f(\mathbf{a}_\alpha)] \equiv  \frac{ \sum\nolimits_{\alpha=1}^{\infty}  \delta (\mathbf{a}_\alpha - \mathbf{a}) f(\mathbf{a}_\alpha)}
      {\sum\nolimits_{\alpha=1}^{\infty}  \delta (\mathbf{a}_\alpha - \mathbf{a})}}
\end{equation}} 
when $\sum_{\alpha=1}^{\infty}\delta (\mathbf{a}_\alpha - \mathbf{a}) \neq 0$, and it is zero otherwise.
Equation (\ref{overall}) states that advective and kinetic correlations alone can not be a net source or sink of probability density when integrated over a closed or infinite domain.
Hence, only after renormalization, the conditional amplitudes $|\psi_\alpha|^2$ can be associated to conditional probabilities, i.e.:
\begin{equation}
 P(\mathbf{R},t|\mathbf{r}^\alpha) = \frac{|\psi_\alpha(\mathbf{R},t)|^2}{\int d\mathbf{R} |\psi_\alpha(\mathbf{R},t)|^2}. 
\end{equation}
Notice that this is different, e.g., from the effective (norm preserving) nuclear wave function considered in the EF framework~\cite{Ali2}.

The derivation of the CD mathematical framework corresponds to the transformation of the many-body TDSE to the \textit{partially} co-moving frame in which the electronic coordinates move attached to the electronic flow 
and the nuclear coordinates are kept in the original inertial frame.
Within the new coordinates, the electronic convective motion is described by a set of trajectories of infinitesimal fluid elements (Lagrangian trajectories), while the nuclear motion is determined by the evolution
of the (nuclear) conditional wave functions in an Eulerian frame. 
The purpose of this partial time-dependent coordinate transformation is to propagate the electronic coordinates along with the electronic flow such that they remain located where the full molecular wave function has a significant amplitude. 
In this way, the amount of ``dead wood'', as it is commonly called in the context of configuration interaction expansions~\cite{deadwood}, can be significantly reduced.
Furthermore, the possibilities of the core ideas behind the CD rely on the fact that the underlying mathematical structure admits approximations which substantially reduce the computational complexity without deteriorating the quality of results.
This has been demonstrated for example in Refs.~\cite{Guil1,JPCL}, where a very simple (zero-order) propagation scheme based on the CD approach was able to accurately describe not only non-adiabatic electronic transitions, 
but also intricate nuclear quantum effects.

\section{On the role of the $\mathbb{C}$-TDPES}\label{BOPES}
To provide some physical intuition on the role played by the $\mathbb{C}$-TDPESs,  we further decompose $V_\alpha$ in Eq.~(\ref{Valpha}) into real and imaginary parts, i.e., 
\begin{eqnarray}
&&V_\alpha^{\Re} = W_\alpha(R,t) + K_\alpha^\Re + A_\alpha^\Re\nonumber,~~\text{and}\nonumber\\ 
&&V_\alpha^{\Im} = K_\alpha^\Im + A_\alpha^\Im,
\end{eqnarray}
where
\begin{eqnarray}\label{kinetic_RI}
  &&K_\alpha^\Re   \!=\!   \sum_{\xi=1}^{N_e} Q^e_{\xi,\alpha}   \!+\!  \frac{(\nabla_\xi S)^2}{2} \Big|_{\mathbf{r}_\alpha}, \label{kinetic_R} \\
  &&K_\alpha^\Im   \!=\!   -\sum_{\xi=1}^{N_e}  \frac{\nabla_\xi \mathbf{j}_\xi}{2|\Psi|^2} \Big|_{\mathbf{r}_\alpha}, \label{kinetic_I}
\end{eqnarray}
and
\begin{eqnarray}\label{advective_RI}
  &&A_\alpha^\Re \!=\! -\sum_{\xi=1}^{N_e} \nabla_\xi S \big|_{\mathbf{r}_\alpha}   \!\cdot\!   \dot\mathbf{r}_{\xi,\alpha} \\
  &&A_\alpha^\Im \!=\!  \sum_{\xi=1}^{N_e}  \frac{\nabla_\xi |\Psi|^2}{2|\Psi|^2} \Big|_{\mathbf{r}_\alpha}   \!\cdot\!   \dot\mathbf{r}_{\xi,\alpha}, 
\end{eqnarray}
are respectively the real and imaginary parts of the kinetic and advective correlation potentials.
In Eq.~(\ref{kinetic_RI}) we have respectively defined the $\xi$-th components of the so-called electronic quantum potential~\cite{BookOriols,AppBohm},
\begin{equation}
 Q^e_{\xi,\alpha} = -\frac{\nabla_\xi^2 |\Psi|}{2|\Psi|}\Big|_{\mathbf{r}_\alpha} 
\end{equation}
and the current probability density $\mathbf{j}_\xi = |\Psi|^2 \nabla_\xi S$. 

While the classical kinetic potential, $(\nabla_\xi S)^2/2$, is in general a smooth function of the nuclear coordinates,  
we expect the contributions, $Q^e_{\xi,\alpha}$ and $K_\alpha^\Im$, to be rather ``discontinuous'' because of their dependence on the inverse of the probability density.
Notice that these last two quantities are implicit functions of the electronic coordinates. 
Specifically, the electronic quantum potential $Q^e_{\xi,\alpha}$ accounts for changes of the curvature of the full probability density along the electronic coordinates (as a function of the nuclear positions). 
Alternatively, $K_\alpha^\Im$ describes the dispersion (e.g. spreading or splitting) of the full wave function in the electronic direction.
Both contributions become large in the vicinity of a node, and thus, they will be important whenever the full probability density splits apart along the electronic coordinates.
The advective correlations in Eq.~(\ref{advective_RI}) are weighted by the electronic velocities, $\dot{\mathbf{r}}_{\xi,\alpha}$, and  
hence, they will be significant during a fast reconfiguration of the electronic degrees of freedom. 

As it will be discussed later, the contribution of the kinetic correlation potential to the full $\mathbb{C}$-TDPESs is much larger than the advective correlation potential in general situations. 
In order to analyse the role played by the kinetic potential in the evolution of the nuclear dynamics we first discuss the connection between the
$\mathbb{C}$-TDPESs and the standard Born-Oppenhaimer (BO) picture. The BO electronic states, $\Phi_\mathbf{R}^{(j)}(\mathbf{r})$, and potential 
energy surfaces, $\epsilon_{BO}^{(j)}(\mathbf{R})$, are defined through:
\begin{equation}\label{kinetic_H}
 \left(\hat T_e + W(\mathbf{r}, \mathbf{R}) \right)\Phi_\mathbf{R}^{(j)}(\mathbf{r}) = \epsilon_{BO}^{(j)}(\mathbf{R}) \Phi_\mathbf{R}^{(j)}(\mathbf{r}).
\end{equation}
If only one BO state, e.g. $\Phi_\mathbf{R}^{(1)}(\mathbf{r})$, is involved in the dynamics, i.e. $ \Psi(\mathbf{r}, \mathbf{R} ,t) = C_1(\mathbf{R} ,t) \Phi_\mathbf{R}^{(1)}(\mathbf{r})$,
then the kinetic correlation potential reads:
\begin{eqnarray}\label{appb_2}
 K_\alpha &=& \epsilon_{BO}^{(1)}(\mathbf{R})  - W_\alpha(\mathbf{R},t).
\end{eqnarray}
In the absence of advective correlations, Eq. (\ref{appb_2}) implies that the $\mathbb{C}$-TDPESs (independently of the electronic trajectory $\mathbf{r}_\alpha$) are all real, time-independent, 
and equal to BOPES $\epsilon_{BO}^{(1)}(\mathbf{R})$, i.e.:
\begin{equation}
 V_\alpha(\mathbf{R},t) = V^\Re_\alpha(\mathbf{R}) = \epsilon_{BO}^{(1)}(\mathbf{R}).
\end{equation}

The situation is, however, not so trivial as we move to a nonadiabatic scenario.
Consider for instance the case where there are only two BO states involved in the dynamics, i.e. $ \Psi(\mathbf{r}, \mathbf{R} ,t) = C_1(\mathbf{R} ,t) \Phi_\mathbf{R}^{(1)}(\mathbf{r}) + C_2(\mathbf{R} ,t) \Phi_\mathbf{R}^{(2)}(\mathbf{r})$.
Then the kinetic correlation potential reads:
\begin{eqnarray}\label{appb_2-b}
 K_\alpha = \epsilon_{BO}^{(1)}(\mathbf{R})  - W_\alpha(\mathbf{R},t) 
                                    +\frac{C_2(\mathbf{R},t)\Delta_{12}(\mathbf{R})\Phi_\mathbf{R}^{(2)}(\mathbf{r})}{
                                    \Psi(\mathbf{r},\mathbf{R},t)}
                                    \Big|_{\mathbf{r}_\alpha} \nonumber\\
          = \epsilon_{BO}^{(2)}(\mathbf{R})  - W_\alpha(\mathbf{R},t) 
                                    -\frac{C_1(\mathbf{R},t)\Delta_{12}(\mathbf{R})\Phi_\mathbf{R}^{(1)}(\mathbf{r})}{
                                    \Psi(\mathbf{r},\mathbf{R},t)}
                                    \Big|_{\mathbf{r}_\alpha}, \quad \;\;
\end{eqnarray}
where we have defined the relation $\epsilon_{BO}^{(2)}(\mathbf{R}) = \epsilon_{BO}^{(1)}(\mathbf{R})+\Delta_{12}(\mathbf{R})$.
By simply separating real and imaginary parts of Eq. (\ref{appb_2}) we can write:
\begin{eqnarray}\label{K_re_bo}
 K_\alpha^\Re  = \epsilon_{BO}^{(1)}(\mathbf{R})  - W_\alpha(\mathbf{R},t)  
                 + \Re\left(\frac{C_2 \Phi_\mathbf{R}^{(2)}(\mathbf{r})}{\Psi(\mathbf{r},\mathbf{R},t)}\right) \Bigg|_{\mathbf{r}_\alpha}\Delta_{12}, \quad   \;\;
\end{eqnarray}
and
\begin{equation}\label{K_im_bo}
 K_\alpha^\Im = \Im\left(\frac{C_2(\mathbf{R},t) \Phi_\mathbf{R}^{(2)}(\mathbf{r})}{\Psi(\mathbf{r},\mathbf{R},t)}\right) \Bigg|_{\mathbf{r}_\alpha}\Delta_{12}(\mathbf{R}). 
\end{equation}
By comparing Eqs. (\ref{kinetic_R}) with (\ref{K_re_bo}) and (\ref{kinetic_I}) with (\ref{K_im_bo}) we can write the following equalities for a two level system (in the absence of advective correlations):
\begin{eqnarray}\label{appb_3_r}
    V_\alpha^\Re &=& \sum_{\xi=1}^{N_e} \Big( Q^e_{\xi,\alpha}  +   \frac{(\nabla_\xi S)^2 |_{\mathbf{r}_\alpha}}{2} \Big) + W_\alpha \nonumber\\
                 &=& \epsilon_{BO}^{(1)} +\Re\left(\frac{C_2 \Phi_\mathbf{R}^{(2)}}{\Psi}\right) \Bigg|_{\mathbf{r}_\alpha}\Delta_{12}, 
\end{eqnarray}
and
\begin{eqnarray}\label{appb_3_i}
    V_\alpha^\Im = - \frac{1}{2}\sum_{\xi=1}^{N_e} \frac{\nabla_\xi \mathbf{j}_\xi}{|\Psi|^2}  \Big|_{\mathbf{r}_\alpha} 
                        = \Im\left(\frac{C_2 \Phi_\mathbf{R}^{(2)}}{\Psi}\right) \Bigg|_{\mathbf{r}_\alpha}\Delta_{12}.
\end{eqnarray}

To better understand the meaning of these quantities, in the following we evaluate and discuss three particularly relevant limits of Eqs. (\ref{appb_3_r}) and (\ref{appb_3_i}): \newline \\
{\bf Case 1}: $|C_2| = 1$ (which implies $|C_1| = 0$ and also $\nabla_\xi S = 0$)
\begin{eqnarray}
  &&V_\alpha^\Re  = \sum_{\xi=1}^{N_e} Q^e_{\xi,\alpha}(\mathbf{R},t) + W_\alpha(\mathbf{R},t) = \epsilon_{BO}^{(2)}(\mathbf{R})  , \nonumber\\
  &&V_\alpha^\Im = -\frac{1}{2}\sum_{\xi=1}^{N_e}  \frac{\nabla_\xi \mathbf{j}_\xi(\mathbf{r},\mathbf{R},t)}{|\Psi(\mathbf{r},\mathbf{R},t)|^2}   \Big|_{\mathbf{r}_\alpha}= 0.
\end{eqnarray}
{\bf Case 2:} $|C_1| = 1$ (which implies $|C_2| = 0$ and also $\nabla_\xi S = 0$)
\begin{eqnarray}
  &&V_\alpha^\Re = \sum_{\xi=1}^{N_e} Q^e_{\xi,\alpha}(\mathbf{R},t) + W_\alpha(\mathbf{R},t) = \epsilon_{BO}^{(1)}(\mathbf{R})  , \nonumber\\
  &&V_\alpha^\Im = -\frac{1}{2} \sum_{\xi=1}^{N_e} \frac{\nabla_\xi \mathbf{j}_\xi(\mathbf{r},\mathbf{R},t)}{|\Psi(\mathbf{r},\mathbf{R},t)|^2}  \Big|_{\mathbf{r}_\alpha}= 0.
\end{eqnarray}
{\bf Case 3:} $\Delta_{12} \rightarrow 0$ (which implies $\nabla_\xi S = 0$)
\begin{eqnarray}
  &&V_\alpha^\Re = \sum_{\xi=1}^{N_e} Q^e_{\xi,\alpha}(\mathbf{R},t) + W_\alpha(\mathbf{R},t) 
                                   = \epsilon_{BO}^{(1\setminus 2)}(\mathbf{R}), \nonumber\\ 
  &&V_\alpha^\Im = -\frac{1}{2} \sum_{\xi=1}^{N_e} \frac{\nabla_\xi \mathbf{j}_\xi(\mathbf{r},\mathbf{R},t)}{|\Psi(\mathbf{r},\mathbf{R},t)|^2} \Big|_{\mathbf{r}_\alpha}= 0.
\end{eqnarray}
Case 1 and Case 2 represent the limits in which one of the BO states is
dominant at a certain time and in a certain region in $\mathbf{R}$-space. 
In these two cases, the imaginary part of the kinetic correlation potential is zero and the real
part is identical with one of the BOPESs. Departing from these limiting cases, the kinetic
potential becomes complex where its real part represents the bridge
between the BO-like pieces of the potential.
Case 3, shows that when the BOPESs come close to each others,
i.e., nearby conical intersections or avoided crossings, the kinetic
potential is real and it coincides with the BOPESs. 
Therefore, in the adiabatic limit as well as in the extreme
non-adiabatic limit the kinetic correlation potential is real.


\section{Numerical Results}\label{numerical}
To study numerically the features of the exact $\mathbb{C}$-TDPESs during nonadiabatic processes in the CD approach, we employ the model introduced by Shin and Metiu~\cite{Metiu}.
This model has become a standard one (just as Tully's models but in full configuration space) as it is simple and provides all the key ingredients to test new approaches to nonadiabatic dynamics. 
The model is very flexible and accepts an infinite number of parameter configurations that give rise to a number of interesting situations where electron-nuclear correlations play a crucial role. 

\begin{figure}
\includegraphics[width=\columnwidth]{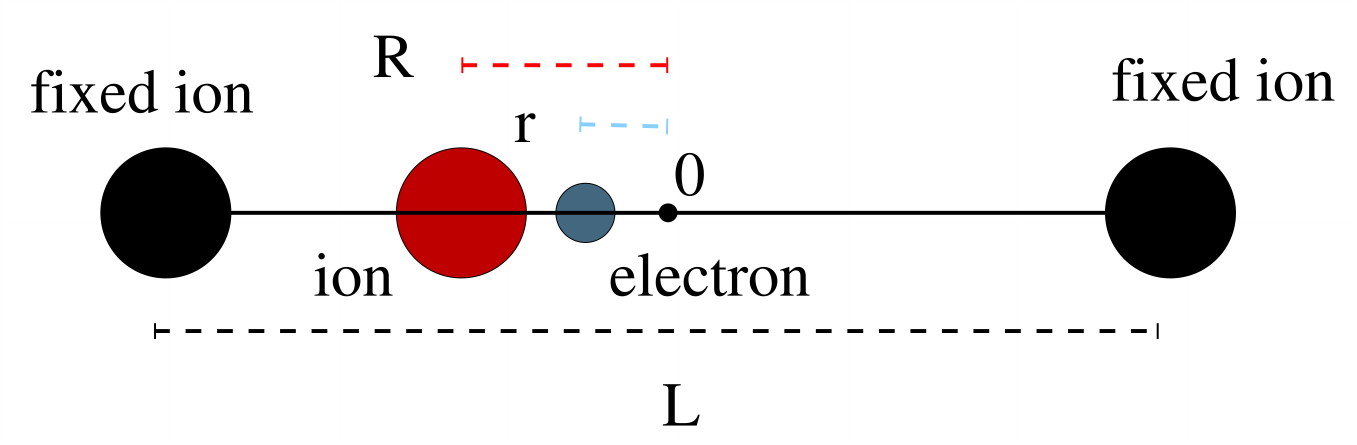}
\caption{Schematic representation of the Shin-Metiu model~\cite{Metiu}. 
Two ions are fixed (in black) and a third one (in red) and an electron (in blue) are free to move in one dimension.}
\label{fig_scheme}
\end{figure}

The Shin-Metiu model consists of three ions and a single electron. 
Two ions are fixed at a distance $L = 19.0a_0$, and the third ion and the electron are free to move in one dimension along the line joining the fixed ions (see Fig.~\ref{fig_scheme}). 
The Hamiltonian for this system reads
\begin{eqnarray}\label{metiu_ham}
\hat H(r,R) = -\frac{1}{2}\frac{\partial^2}{\partial r^2} - \frac{1}{2M}\frac{\partial^2}{\partial R^2} + \frac{1}{|\frac{L}{2} - R|} + \frac{1}{|\frac{L}{2} + R|} \nonumber\\
- \frac{\text{erf}\big(\frac{|R - r|}{R_f}\big)}{|R - r|}  - \frac{\text{erf}\big(\frac{|r - \frac{L}{2}|}{R_r}\big)}{|r - \frac{L}{2}|}  - \frac{\text{erf}\big(\frac{|r + \frac{L}{2}|}{R_l}\big)}{|r + \frac{L}{2}|}, \quad
\end{eqnarray}
where $erf()$ represents the error function, the symbols $\mathbf{r}$ and $\mathbf{R}$ are replaced by $r$ and $R$, and the coordinates of the electron and the movable nucleus are measured from the center of the two fixed ions. 
In this work, we choose the remaining parameters to be $M = 1836$a.u. and $R_f = 5.0a_0$, $R_l = 4.0a_0$, and $R_r = 3.1a_0$ such that the first BOPES, $\epsilon^{(1)}_{BO}$, 
is strongly coupled to the second, $\epsilon^{(2)}_{BO}$ around $R_{ac} = -2a_0$. The coupling to the remaining BOPESs is negligible. 

Let us remark that by taking such a confined system, one is exposed to strong electron-nuclear correlations that usually do not show up in scattering problems with a continuum of states. 
It is specially under this highly confined conditions that the existing (nonadiabatic) theories fail to describe the correlated motion of electrons and nuclei.
Notice also that the parameters used in this work differ significantly from those used in Ref.~\cite{Guil1}. 
More specifically, in Ref.~\cite{Guil1} we described nuclear quantum effects such as interferences or tunneling, while in this work we delve into the splitting of the nuclear probability density. 

\begin{figure}
\includegraphics[width=\columnwidth]{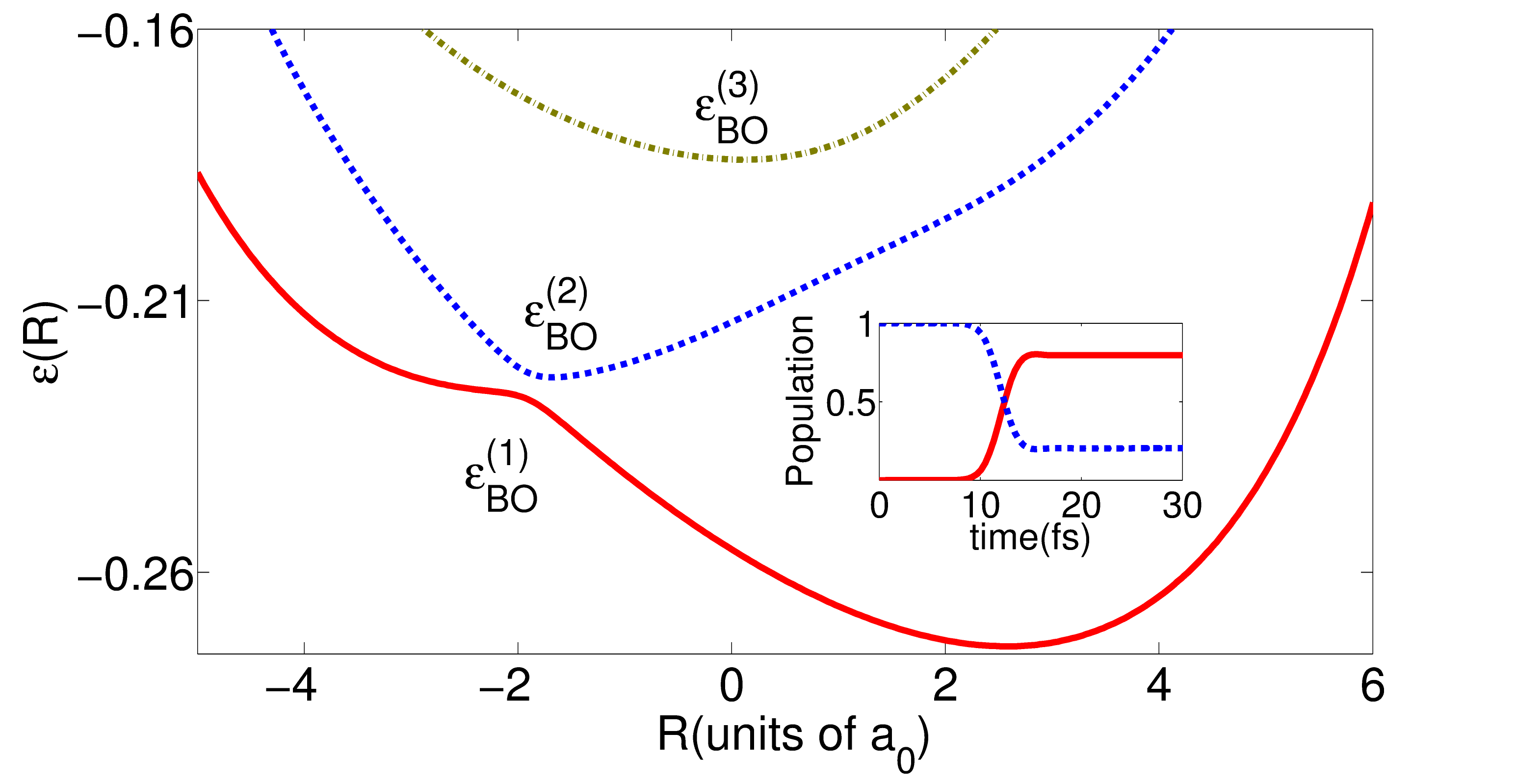}
\caption{Lowest three BOPESs $\epsilon_{BO}^{(1)}$, $\epsilon_{BO}^{(2)}$, and $\epsilon_{BO}^{(3)}$. In the inset: adiabatic populations as a function of time.}
\label{fig_1}
\end{figure}

\begin{figure}
\includegraphics[width=\columnwidth]{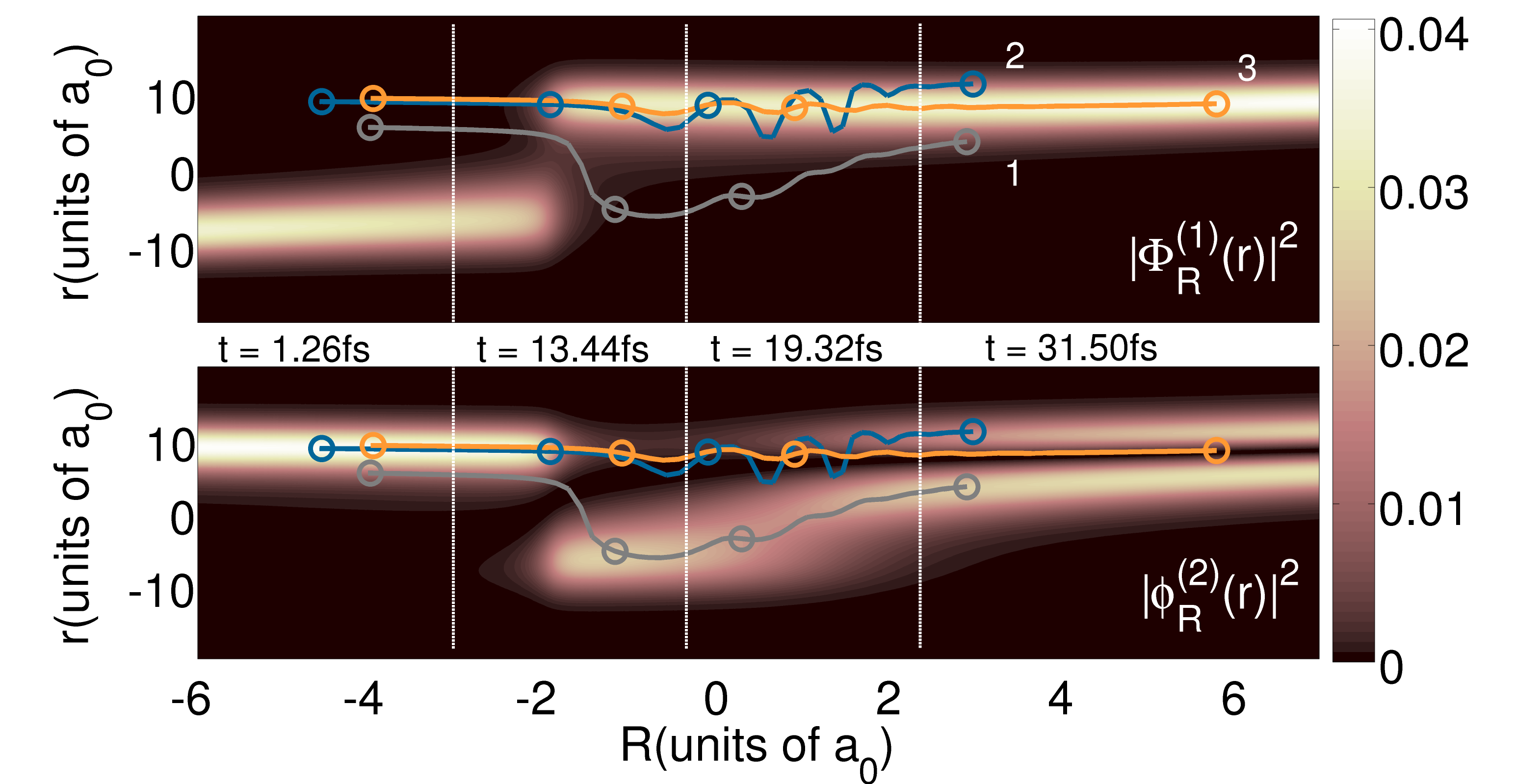}
\caption{Squared value of the Born-Oppenheimer states $\Phi_R^{(1)}(r)$ and $\Phi_R^{(2)}(r)$ along with the evolution of three selected trajectories $\{r_\alpha(t),R_\alpha(t)\}$ labeled $\alpha=1,2,3$. 
The position of these trajectories at four different times is also shown for later reference.}
\label{fig_2}
\end{figure}

\begin{figure*}
\includegraphics[width=\textwidth,height=5.5cm]{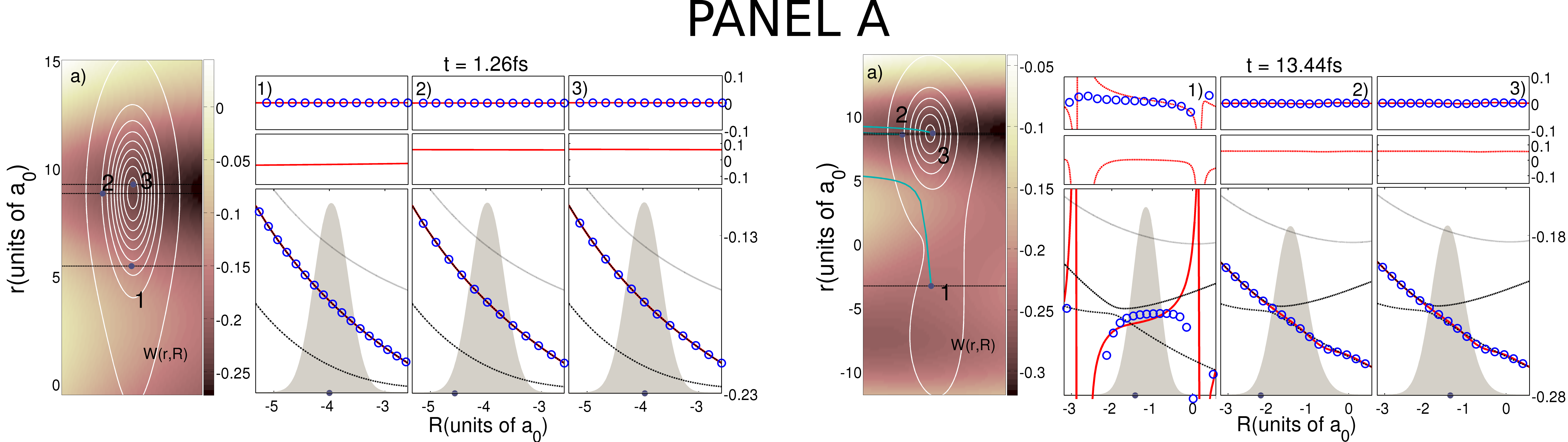}
\caption{
The nuclear dynamics before the splitting of the probability density is mainly characterized by $\mathbb{C}$-TDPESs smoothly connecting $\epsilon_{BO}^{(2)}$ and $\epsilon_{BO}^{(1)}$.  
PANEL a: Contour lines of the full electron-nuclear probability density (in white) together with three conditional wave functions , $\psi_1(R,t)$, $\psi_2(R,t)$, and $\psi_3(R,t)$ (black lines), defined along with trajectories $\{r_{\alpha=1,2,3},R_{\alpha=1,2,3}\}$ (in cyan). 
In the background (in copper color scale), full (2D) dependence of the electron-nuclear potential energy $W(r,R)$.
PANELS 1,2,3 (bottom): Conditional amplitudes $|\psi_{\alpha=1,2,3}(R,t)|^2$ (filled in gray) along with the first three BOPESs (in black) and the real part of the $\mathbb{C}$-TDPESs $V_\alpha^\Re = W_\alpha + K^\Re_\alpha + A^\Re_\alpha$ (in red). 
For comparison we also show the potential $V_{\alpha,app}^\Re = Q_\alpha^e + W_\alpha$ (blue circles).
PANELS 1,2,3 (middle): Electronic quantum potential $Q_\alpha^e$.
PANELS 1,2,3 (top): Imaginary part of the $\mathbb{C}$-TDPESs $V_\alpha^\Im = K^\Im_\alpha + A^\Im_\alpha$ (red line) along with the approximated potential $V_{\alpha,app}^\Im = K_\alpha^\Im$ (blue circles).
}
\label{fig_3}
\end{figure*}

The time-independent electronic problem is fully characterized through the so-called electronic Hamiltonian, i.e. $(\hat T_e + W)\Phi_{R}^{(j)}(r) = \epsilon^{(j)}_{BO}(R) \Phi_{R}^{(j)}(r)$.
We suppose the system to be initially excited to $\epsilon^{(2)}_{BO}$ and the initial nuclear wave function to be a Gaussian wavepacket with  
$\sigma = 1/\sqrt{2.85}$, centered at $R = -4.0a_0$. 
On Fig.~\ref{fig_1} we show the first three BOPESs together with the evolution of the adiabatic populations (in the inset).
Figure~\ref{fig_2} shows the first two Born-Oppenheimer states $\Phi_R^{(1)}(r)$ and $\Phi_R^{(2)}(r)$ along with the evolution of three selected trajectories $\{r_\alpha,R_\alpha\}$ 
labeled $\alpha=1,2,3$ that will be used to analyse the $\mathbb{C}$-TDPESs. In Fig.~\ref{fig_3} we present four time snapshots containing relevant information about the $\mathbb{C}$-TDPESs as well as the conditional nuclear wave functions.
For the sake of clarity we also define approximated real and imaginary components of $\mathbb{C}$-TDPESs respectively as $V_{\alpha,app}^\Re = Q_\alpha^e + W_\alpha$ and 
$V_{\alpha,app}^\Im = K_\alpha^\Im$.

\begin{figure*}
\includegraphics[width=\textwidth,height=5.5cm]{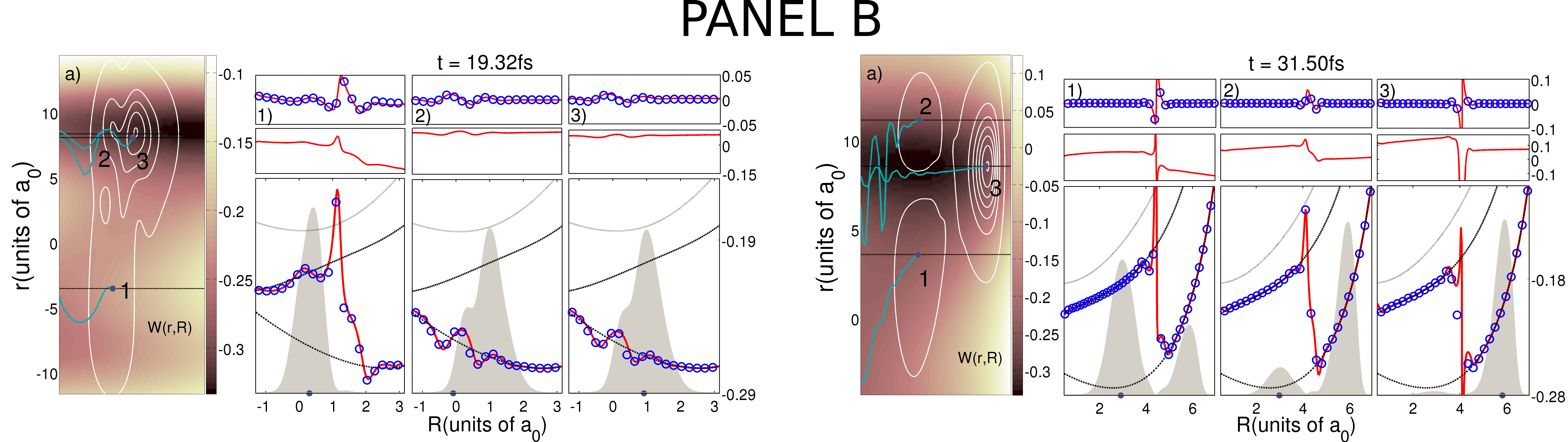}
\caption{
The nuclear dynamics during the splitting of the probability density is accompanied by discontinuous steps in the $\mathbb{C}$-TDPESs piecewise connecting $\epsilon_{BO}^{(2)}$ and $\epsilon_{BO}^{(1)}$.   
More detaails can be found in the caption of Fig.~\ref{fig_3}.
}
\label{fig_4}
\end{figure*}

At the initial time, due to the choice of $\Psi(r,R,t_0)$, the $\mathbb{C}$-TDPESs are real, 
$\alpha$-independent, and by construction all equal to the first excited BOPES $\epsilon_{BO}^{(2)}$: 
\begin{equation}
 V_\alpha = \frac{\hat T_e \Phi^{(2)}_R(r)}{\Phi^{(2)}_R(r)}\Bigg|_{r_\alpha} + W_\alpha = V_{\alpha,app}^\Re = \epsilon_{BO}^{(2)}, \quad \forall \alpha
\end{equation}
where we used that $Q_\alpha^e(R,t_0) = {\hat T_e \Phi^{(2)}_R(r)}/{\Phi^{(2)}_R(r)}$ when $\partial_r S = 0$.
As it is made clear in the right panels of Fig.~\ref{fig_2}, at $t = 13.44$fs, the trajectories $(r_2,R_2)$ and $(r_3,R_3)$, associated respectively with the conditional wave functions  $\psi_2$ and $\psi_3$, 
are running straight (i.e. $\dot{r}_{2,3} \approx 0$) from the support of $\Phi_R^{(2)}(r)$ to fall in the support of $\Phi_R^{(1)}(r)$. 
For $\alpha=2,3$ the $\mathbb{C}$-TDPESs are real ($V_{2,3}^\Im = 0$) and satisfy 
$V_{2,3} \approx V_{2,3,app}^\Re$, with $Q_{2,3}^e(R<R_{ac}) \approx {\hat T_e \Phi^{(2)}_R(r)}/{\Phi^{(2)}_R(r)}$ and $Q_{2,3}^e(R>R_{ac}) \approx {\hat T_e \Phi^{(1)}_R(r)}/{\Phi^{(1)}_R(r)}$.
Therefore, as can be seen from Fig.~\ref{fig_3} (bottom panels), the $\mathbb{C}$-TDPESs resemble diabatic potential-energy surfaces, coinciding with $\epsilon_{BO}^{(2)}$ for $R<R_{ac}$ and smoothly going through 
the avoided crossing region to follow $\epsilon_{BO}^{(1)}$ for $R>R_{ac}$.
The trajectory $(r_1,R_1)$, while staying most of the time in the support of $\Phi_R^{(2)}(r)$, is now tunneling from one Coulomb potential well to the other.
The gradient of the phase, $(\nabla_r S)|_{r_1}$, that appears in both the kinetic and advective correlation terms (also implicitly in the electronic velocity $\dot{r}_1$), 
is now very large due to the fast reconfiguration of the electronic degrees of freedom. Hence $V_{1,app}^\Re$ and $V_1$ show important differences.
%
%
At the later time $t = 19.32 $fs, the advective correlation and classical kinetic terms become once more negligible, and therefore $V_\alpha \approx V_{\alpha,app}^\Re + V_{\alpha,app}^\Im$ (see Fig.~\ref{fig_4}).
While $(r_1,R_1)$ stays preferentially in the support of $\Phi^{(2)}_R(r)$, the trajectories $(r_2,R_2)$ and $(r_3,R_3)$ are sampling the support of $\Phi^{(1)}_R(r)$ and $\Phi^{(2)}_R(r)$ simultaneously (see Fig.~\ref{fig_2}).
The conditional wave functions  $\psi_1$, $\psi_2$ and $\psi_3$ are now linear combinations of $\Phi^{(2)}_R(r)$ and $\Phi^{(1)}_R(r)$. 
This mixture leads to the formation of the step observed for $V_1$ around $R=1.5a_0$, and two wiggles accompanying the $\mathbb{C}$-TDPESs $V_{2}$ and $V_{3}$. 
All these features indicate the nonadiabatic nature of the conditional wave functions  and can be directly associated with the formation of nodes in the full probability density.
Finally, at time $t = 31.5$ fs, the full molecular wave function has been split both along the electronic and nuclear directions.  
While the three conditional wave functions , $\psi_1$, $\psi_2$ and $\psi_3$, still embody contributions from $\Phi_R^{(1)}(r)$ and $\Phi_R^{(2)}(r)$, these contributions are now very well separated along the nuclear coordinates 
with a minimum at around $R_{sp} = 4a_0$ (see also Fig.~\ref{fig_2}). 
For nuclear coordinates less than $R_{sp}$, the support of the full probability density perfectly fits in the support of the first excited state $\Phi_R^{(2)}(r)$,
while for  $R > R_{sp}$ it mainly falls in the ground state $\Phi_R^{(1)}(r)$. 
As a direct consequence, the quantum electronic potential acquires a discontinuity, i.e.
$Q_\alpha^e(R<R_{sp}) \approx {\hat T_e \Phi^{(2)}_R(r)}/{\Phi^{(2)}_R(r)}$ while $Q_\alpha^e(R>R_{sp}) \approx {\hat T_e \Phi^{(1)}_R(r)}/{\Phi^{(1)}_R(r)}$.
The real parts of the $\mathbb{C}$-TDPESs are therefore piecewise connecting adiabatic surfaces, i.e. 
$V_\alpha^\Re  \approx V_{\alpha,app}^\Re \approx \epsilon_{BO}^{(1)}$ for $R>R_{sp}$, and $V_\alpha^\Re  \approx V_{\alpha,app}^\Re \approx \epsilon_{BO}^{(2)}$ for $R<R_{sp}$. 
In the transition between these two regions (at around $R_{sp}$), the conditional wave functions  $\psi_1$, $\psi_2$ and $\psi_3$ become linear combinations of the lowest two adiabatic states.
As a result, the $\mathbb{C}$-TDPESs exhibit abrupt peaks, for both real and imaginary parts, mainly originating from the potentials $Q_\alpha^e$ and $K_\alpha^\Im$.

\section{Discussion}\label{discussion}
In view of the above discussion, the characteristic features of an ensemble of $\mathbb{C}$-TDPESs of the CD framework during a nonadiabatic transition are very similar to that of the gauge invariant part of the
exact TDPES~\cite{Ali_JCP,Ali2,AASMG2015} within the EF framework: while in the vicinity of the avoided crossing both potentials follow a diabatic shape, far from the region of avoided crossings 
(when the nonadiabatic couplings are negligible) they are piecewise identical with two different adiabatic BOPESs with nearly discontinuous steps in between. By discussing 
the link between the potentials of the EF and CD approaches, in the remaining part of the manuscript we elaborate on the origin of the discontinuities, wich 
seem to be a universal feature of the time-dependent potentials driving the nuclear dynamics within formally exact approaches to correlated electron-nuclear dynamics. 

The CD and the EF approaches are essentially different exact mathematical formulations of the electron-nuclear coupled problem. 
While the CD approach is a ``trajectory-guided'' method that treats electrons and nuclei on an equal mathematical footing~\cite{Guil1}, the main equations of the EF approach arise from Frenkel's stationary action principle and
hence involve the tracing out of either the electronic or nuclear degrees of freedom~\cite{Ali1}. 
Therefore, while the CD and the EF approaches can be formally linked together (as it will be briefly outlined in the following), there is no straightforward connection between the conditional wave function $\psi_\alpha(\mathbf{R},t)$ of the CD 
and the effective wave functions  of the EF approach.

The exact TDPES that governs the  nuclear dynamics together with a vector potential within the EF framework consists of two parts
~\cite{Ali_JCP,Ali2,AASMG2015}, the gauge invariant term, $\epsilon_{gi}(\mathbf{R},t)$, and a gauge dependent term (see Appendix A for a full derivation of the connection between the CD and the EF approaches). By virtue of the EF 
theorem, the conditional nuclear wave function can be written as a direct product of electronic and nuclear wave functions , i.e., $\psi_\alpha(\mathbf{R},t) = \Phi_\mathbf{R}(\mathbf{r}_\alpha,t) \chi(\mathbf{R},t)$. 
Hence, $\epsilon_{gi}(\mathbf{R},t)$ can be expressed in terms of $\mathbb{C}$-TDPESs and conditional nuclear wave functions , $\psi_\alpha(\mathbf{R},t)$, as:
\begin{equation}
\label{connection}
 \epsilon_{gi}(\mathbf{R},t) = \frac{\int {\mathcal{D}}_\mathbf{r}[|\psi_\alpha|^2 V_\alpha]d\mathbf{r}}{\int {\mathcal{D}}_\mathbf{r}[|\psi_\alpha|^2]d\mathbf{r}}+ O(M_n^{-1}). 
\end{equation}
As the second term, $O(M_n^{-1})$, is mostly negligible due to its dependence on the inverse of the nuclear mass, the first term on the r.h.s of Eq.~(\ref{connection}) establishes a direct connection between 
the gauge invariant part of the TDPES and the $\mathbb{C}$-TDPESs: $\epsilon_{gi}(\mathbf{R},t)$ is an ensemble average of $\mathbb{C}$-TDPESs, $V_\alpha$, which are integrated along 
the conditional amplitudes $|\psi_\alpha|^2$ and are weighted afterwards by the nuclear probability density, 
$\int{\mathcal{D}}_\mathbf{r}[|\psi_\alpha|^2]d\mathbf{r} = \int |\Psi|^2 d\mathbf{r}$. As $\int {\mathcal{D}}_\mathbf{r}[|\psi_\alpha|^2 V^\Im_\alpha]d\mathbf{r}$ is zero (see in Appendix A the full derivation), 
Eq.~(\ref{connection}) can be written as:
\begin{equation}
 \epsilon_{gi}(\mathbf{R},t) \approx \frac{\int {\mathcal{D}}_\mathbf{r}[|\psi_\alpha|^2 V^\Re_\alpha]d\mathbf{r}}{\int {\mathcal{D}}_\mathbf{r}[|\psi_\alpha|^2]d\mathbf{r}}.
\end{equation}
Furthermore, the electronic quantum potentials are the main source of the discontinuities of $V^\Re_\alpha$ as discussed before. Hence, the steps in $\epsilon_{gi}(\mathbf{R},t)$ can be viewed as 
an ensemble average of the discontinuities of the electronic quantum potential.

\section{Conclusion}\label{conclusion}
The branching of nuclear wavepackets at nonadiabatic transitions is a key nuclear quantum effect in photophysics and it is traditionally described within the Born-Oppenheimer picture
where the dynamics is described by means of population transfer among different BOPESs.
Within the exact factorization and conditional decomposition frameworks the same physics can also be successfully described by a single effective time-dependent potential that drives the nuclear dynamics along adiabatic segments of BOPESs connected by sudden transitions or jumps. 
Within the CD framework, an ensemble of $\mathbb{C}$-TDPESs governs the dynamics of the conditional nuclear wave functions  and provide us with an alternative approach for the study of nonadiabatic processes. 
In particular, we demonstrated that within the EF framework the discontinuous steps that connects two adiabatic  pieces of the TDPES can be reproduced by an ensemble average of discontinuities of the $\mathbb{C}$-TDPESs.

Our study reveals that the shape and the position of the steps are mainly dictated (both in the CD and the EF approaches) by the electronic quantum potential, $Q^e_\alpha$,
which is implicit in the electronic kinetic energy operator of the full time-dependent Schr\"odinger equation. 
The discontinuities associated to the quantum electronic potential are inevitable when a closed time-dependent form for the electron-nuclear wave function is used (as in the EF, CD or any other equivalent representation) and give rise to a universal mechanism for 
the explanation of the branching of the nuclear probability density in the nonadiabatic regime (beyond the Born-Oppenheimer picture). 
In other words, the paradigm shift associated with the transition from the many BOPESs to the single time-dependent potential entails a discontinuous behavior of the resulting effective time-dependent potentials. 
Moreover, these discontinuities represent a clear signature of the ultimate quantum nature of electron-nuclear correlations; the effective (nuclear) potentials have a direct dependence 
on the full wave function through the action of the electronic quantum potential, which is ultimately the source of the discontinuities of the time-dependent surfaces within the CD and EF frameworks.
 
It is worth mentioning that other attempts to reduce the complexity of the time-dependent (many-electron/ion) Schr\"odinger equation, such as in time-dependent density-functional theory, 
also lead to the emergence of striking discontinuities in the effective time-dependent potentials that govern the reduced variables of interest. 
This is the case, e.g., in the field of electron-electron correlated dynamics, where the nature of such discontinuities is currently under study~\cite{elliott2012universal,luo2013absence,luo2014kinetic}.

\textbf{Acknowledgements}
G.A acknowledges financial support from the Beatriu de Pin\'os program through the Project 2014-BP-B-00244 and the Spanish grant CTQ2016-76423-P.
I.T acknowledges the Swiss National Science Foundation (SNF) through Grant No. 200021-146396 and NCCR-MARVEL.
We also acknowledge financial support from the European Research Council (ERC-2015-AdG-694097), Grupos Consolidados (IT578-13), COST Actions CM1204 (XLIC) and MP1306 (EUSpec), and 
the European Union's H2020  programme under Grant Agreement No. 676580 (NOMAD).
A.A. acknowledges funding from the European Union Horizon 2020 research and innovation programme under the Marie Sklodowska-Curie Grant Agreement No. 702406.

\section*{Appendix A: Derivation of Equation (\ref{connection})}\label{appendix}

In order to establish a formal connection between the $\mathbb{C}$-TDPESs and the TDPES concepts we first write the molecular wave function as a direct product of electronic and nuclear probability amplitudes, i.e.
\begin{equation}\label{cond_factor}
 \psi_\alpha(\mathbf{R},t) = \Phi_\mathbf{R}(\mathbf{r}_\alpha,t) \chi(\mathbf{R},t).
\end{equation} 
Using the transformation ${\mathcal{D}}_\mathbf{r}$ together with the definition of the nuclear probability density, $|\chi(\mathbf{R},t)|^2 = {\int d\mathbf{r} |\Psi(\mathbf{r},\mathbf{R},t)|^2}$, 
one gets the following partial normalization condition,
\begin{equation}\label{partial_norm}
  \int {\mathcal{D}}_{\mathbf{r}}\left[|\Phi_{\mathbf{R}}(\mathbf{r}_\alpha,t)|^2\right] d\mathbf{r} = 1.
\end{equation}
Introducing the above definition in (\ref{cond_factor}) into Eq. (1) we find:
\begin{widetext}
\begin{eqnarray}\label{der1}
i d_t \Phi_{\mathbf{R}}(\mathbf{r}_\alpha,t)  \chi(\mathbf{R},t) + i\Phi_{\mathbf{R}}(\mathbf{r}_\alpha,t) d_t\chi(\mathbf{R},t) = 
    \chi(\mathbf{R},t) \hat T_n \Phi_{\mathbf{R}}(\mathbf{r}_\alpha,t) + \Phi_{\mathbf{R}}(\mathbf{r}_\alpha,t) \hat T_n \chi(\mathbf{R},t) \nonumber\\
  - \sum\limits_{\nu=1}^{N_n}\frac{1}{M_\nu} (\nabla_\nu \Phi_{\mathbf{R}}(\mathbf{r}_\alpha,t)) \nabla_\nu \chi(\mathbf{R},t) 
  + \hat W(\mathbf{r}_\alpha,\mathbf{R},t) \Phi_{\mathbf{R}}(\mathbf{r}_\alpha(t),t) \chi(\mathbf{R},t) \nonumber\\
  + \hat T_e \Phi_{\mathbf{R}}(\mathbf{r},t)\big{|}_{\mathbf{r}_\alpha} \chi(\mathbf{R},t) 
  + i \sum\limits_{\xi=1}^{N_e} \nabla_\xi  \Phi_{\mathbf{R}}(\mathbf{r},t)\big{|}_{\mathbf{r}_\alpha} \cdot \dot{\mathbf{r}}_{\xi,\alpha} \chi(\mathbf{R},t). 
\end{eqnarray}
Multiplying Eq. (\ref{der1}) from the left by $\Phi^*_{\mathbf{R}}(\mathbf{r}_\alpha,t)$, applying the transformation ${\mathcal{D}}_\mathbf{r}$ to the resulting equation, and finally integrating over 
the electronic coordinates, an equation of motion for the nuclear probability density can be isolated:
\begin{eqnarray}\label{der_1}
i d_t\chi(\mathbf{R},t) = \Bigg\{  \hat T_n  
  + \int {\mathcal{D}}_\mathbf{r}[\Phi^*_{\mathbf{R}}(\mathbf{r}_\alpha,t) \hat T_n \Phi_{\mathbf{R}}(\mathbf{r}_\alpha,t)]d\mathbf{r} 
  - \sum\limits_{\nu=1}^{N_n}\frac{1}{M_\nu}\frac{\nabla_\nu \chi(\mathbf{R},t)}{\chi(\mathbf{R},t)} \int {\mathcal{D}}_\mathbf{r}[\Phi^*_{\mathbf{R}}(\mathbf{r}_\alpha,t) \nabla_\nu \Phi_{\mathbf{R}}(\mathbf{r}_\alpha,t)]d\mathbf{r}  \nonumber\\
  + \int {\mathcal{D}}_\mathbf{r}[\Phi^*_{\mathbf{R}}(\mathbf{r}_\alpha,t) \hat W(\mathbf{r}_\alpha,\mathbf{R},t) \Phi_{\mathbf{R}}(\mathbf{r}_\alpha,t)]d\mathbf{r} 
  + \int {\mathcal{D}}_\mathbf{r}[\Phi^*_{\mathbf{R}}(\mathbf{r}_\alpha,t) \hat T_e \Phi_{\mathbf{R}}(\mathbf{r},t)\big{|}_{\mathbf{r}_\alpha}]d\mathbf{r}  \nonumber\\
  + i \int {\mathcal{D}}_\mathbf{r}[\Phi^*_{\mathbf{R}}(\mathbf{r}_\alpha,t) \sum\limits_{\xi=1}^{N_e} \nabla_\xi  \Phi_{\mathbf{R}}(\mathbf{r},t) \big{|}_{\mathbf{r}_\alpha} \cdot \dot{\mathbf{r}}_{\xi,\alpha} ]d\mathbf{r} 
  + i \int {\mathcal{D}}_\mathbf{r}[\Phi^*_{\mathbf{R}}(\mathbf{r}_\alpha,t) d_t \Phi_{\mathbf{R}}(\mathbf{r}_\alpha,t)]d\mathbf{r}  
\Bigg\} \chi(\mathbf{R},t),
\end{eqnarray}
\end{widetext}
Notice that in Eq.~(\ref{der_1}) we have used the partial normalization condition defined in (\ref{partial_norm}).
Using the hydrodynamic derivative, $d_t \Phi_{\mathbf{R}}(\mathbf{r}_\alpha,t)  =  \partial_t \Phi_{\mathbf{R}}(\mathbf{r}_\alpha,t)  + \sum_\xi^{N_e} \nabla_\xi  \Phi_{\mathbf{R}}(\mathbf{r},t)  |_{\mathbf{r}_\alpha} \cdot \dot{\mathbf{r}}_{\xi,\alpha}$, 
Eq. (\ref{der_1}) can be rewritten as:
\begin{widetext}
\begin{eqnarray}\label{last_step}
i d_t\chi(\mathbf{R},t) = \Bigg\{  \hat T_n  
  + \int {\mathcal{D}}_\mathbf{r}[\Phi^*_{\mathbf{R}}(\mathbf{r}_\alpha,t) \hat T_n \Phi_{\mathbf{R}}(\mathbf{r}_\alpha,t)]d\mathbf{r} 
  - \sum\limits_{\nu=1}^{N_n}\frac{1}{M_\nu}\frac{\nabla_\nu \chi(\mathbf{R},t)}{\chi(\mathbf{R},t)} \int {\mathcal{D}}_\mathbf{r}[\Phi^*_{\mathbf{R}}(\mathbf{r}_\alpha,t) \nabla_\nu \Phi_{\mathbf{R}}(\mathbf{r}_\alpha,t)]d\mathbf{r}  \nonumber\\
  + \int {\mathcal{D}}_\mathbf{r}[\Phi^*_{\mathbf{R}}(\mathbf{r}_\alpha,t) \hat W_\alpha \Phi_{\mathbf{R}}(\mathbf{r}_\alpha,t)]d\mathbf{r}  
  + \int {\mathcal{D}}_\mathbf{r}[\Phi^*_{\mathbf{R}}(\mathbf{r}_\alpha,t) \hat T_e \Phi_{\mathbf{R}}(\mathbf{r},t)\big{|}_{\mathbf{r}_\alpha}]d\mathbf{r}  
  + i \int {\mathcal{D}}_\mathbf{r}[\Phi^*_{\mathbf{R}}(\mathbf{r}_\alpha,t) \partial_t \Phi_{\mathbf{R}}(\mathbf{r}_\alpha,t)]d\mathbf{r}
\Bigg\} \chi(\mathbf{R},t).  \qquad
\end{eqnarray}
\end{widetext}
We can now identify $\mathbf{\mathcal{A}}_\nu(\mathbf{R},t) = \int {\mathcal{D}}_\mathbf{r}[\Phi^*_{\mathbf{R}}(\mathbf{r}_\alpha,t) \nabla_\nu \Phi_{\mathbf{R}}(\mathbf{r}_\alpha,t)]d\mathbf{r}$ 
with the $\nu-$component of the time-dependent Berry phase~\cite{Ali1}. 
By adding and removing $\sum\nolimits_{\nu=1}^{N_n} \mathbf{\mathcal{A}}_\nu^2/M_\nu \cdot \chi$, $\sum\nolimits_{\nu=1}^{N_n} i\nabla_\nu \mathbf{\mathcal{A}}_\nu/2M_\nu \cdot \chi$, and 
$\sum\nolimits_{\nu=1}^{N_n} i \mathbf{\mathcal{A}}_\nu/M_\nu \cdot \nabla_\nu \chi$, Eq. (\ref{last_step}) can be finally written in a compact form as
\begin{equation}\label{cond_nuc}
i\partial_t \chi(\mathbf{R},t)   =   \Bigg\{  \sum\limits_{\nu=1}^{N_n}\frac{\big(  -i\nabla_\nu  +  \mathbf{\mathcal{A}}_\nu  \big)^2}{2M_\nu}  
    +   \epsilon(\mathbf{R},t) \Bigg\}\chi(\mathbf{R},t),
\end{equation}
where the TDPES $\epsilon(\mathbf{R},t)$ are defined here in terms of conditional electronic amplitudes $\Phi_{\mathbf{R}}(\mathbf{r}_\alpha,t)$ as:
\begin{widetext}
\begin{eqnarray}\label{TDPES}
\epsilon(\mathbf{R},t) =  
  \int {\mathcal{D}}_\mathbf{r}[\Phi^*_{\mathbf{R}}(\mathbf{r}_\alpha,t) \hat T_e \Phi_{\mathbf{R}}(\mathbf{r},t)\big{|}_{\mathbf{r}_\alpha}]d\mathbf{r}  
  +  \int {\mathcal{D}}_\mathbf{r}[\Phi^*_{\mathbf{R}}(\mathbf{r}_\alpha,t) W_\alpha \Phi_{\mathbf{R}}(\mathbf{r}_\alpha,t)] d\mathbf{r} 
  +  \int {\mathcal{D}}_\mathbf{r}[\Phi^*_{\mathbf{R}}(\mathbf{r}_\alpha,t)  \hat T_n  \Phi_{\mathbf{R}}(\mathbf{r}_\alpha,t)] d\mathbf{r}  \nonumber \\ 
  +  \int {\mathcal{D}}_\mathbf{r}[\Phi^*_{\mathbf{R}}(\mathbf{r}_\alpha,t) \left(\sum\limits_{\nu=1}^{N_n}\frac{1}{2M_\nu} \left( \nabla_{\nu} \mathbf{\mathcal{A}}_\nu - \mathbf{\mathcal{A}}_\nu^2 \right) \right)  \Phi_{\mathbf{R}}(\mathbf{r}_\alpha,t)] d\mathbf{r} 
  -i  \int {\mathcal{D}}_\mathbf{r}[\Phi^*_{\mathbf{R}}(\mathbf{r}_\alpha,t) \partial_t  \Phi_{\mathbf{R}}(\mathbf{r}_\alpha,t)] d\mathbf{r}.  \quad
\end{eqnarray}
\end{widetext}
The above expression for the TDPES can be greatly simplified by gathering all terms that depend on the inverse of the nuclear mass into $O(M_n^{-1})$
\begin{eqnarray}\label{TDPES_2}
\epsilon(\mathbf{R},t) =  
  \int {\mathcal{D}}_\mathbf{r}[\Phi^*_{\mathbf{R}}(\mathbf{r}_\alpha,t) (\hat T_e + W) \Phi_{\mathbf{R}}(\mathbf{r},t) |_{\mathbf{r}_\alpha}]d\mathbf{r}  \nonumber\\
  +  \int {\mathcal{D}}_\mathbf{r}[\Phi^*_{\mathbf{R}}(\mathbf{r}_\alpha,t) (- i\partial_t) \Phi_{\mathbf{R}}(\mathbf{r}_\alpha,t)] d\mathbf{r} 
  + O(M_n^{-1}).  
\end{eqnarray}
We can now use the definition of $\Phi_{\mathbf{R}}(\mathbf{r}_\alpha,t)$ in Eq.~(\ref{cond_factor}) to rewrite the above equation in terms of conditional nuclear wave functions  as:
\begin{widetext}
\begin{eqnarray}\label{TDPES_3}
 \epsilon(\mathbf{R},t) = 
   \frac{1}{|\chi(\mathbf{R},t)|^2} \int {\mathcal{D}}_\mathbf{r}\Big[|\psi_\alpha(\mathbf{R},t)|^2 \Big(  W_\alpha(\mathbf{R},t) 
                      + \frac{\hat T_e \Psi(\mathbf{r},\mathbf{R},t)}{\Psi(\mathbf{r},\mathbf{R},t)}\Big|_{\mathbf{r}_\alpha}  
                      +i\sum_\xi^{N_e} \frac{\nabla_\xi \Psi}{\Psi} \Big|_{\mathbf{r}_\alpha} \cdot \dot{\mathbf{r}}_{\alpha,\xi} \Big)   \Big]d\mathbf{r} \nonumber\\
  + \int {\mathcal{D}}_\mathbf{r} [ \Phi^*_\mathbf{R}(\mathbf{r}_\alpha,t) (-i d_t) \Phi_\mathbf{R}(\mathbf{r}_\alpha,t) ]
  + O(M_n^{-1}).
\end{eqnarray}
\end{widetext}
Using that $|\chi(\mathbf{R},t)|^2 = \int {\mathcal{D}}_\mathbf{r}[|\psi_\alpha(\mathbf{R},t)|^2]d\mathbf{r}$ and identifying the exact $\mathbb{C}$-TDPESs, $V_\alpha$, with the integrand of the first term on the r.h.s of 
Eq. (\ref{TDPES_3}), we can finally write
\begin{eqnarray}
 \epsilon(\mathbf{R},t) = \frac{\int {\mathcal{D}}_\mathbf{r}[|\psi_\alpha(\mathbf{R},t)|^2 V_\alpha   ]d\mathbf{r}}{\int {\mathcal{D}}_\mathbf{r}[|\psi_\alpha(\mathbf{R},t)|^2]d\mathbf{r}} \nonumber\\
    + \int {\mathcal{D}}_\mathbf{r} [ \Phi^*_\mathbf{R}(\mathbf{r}_\alpha,t) (-i d_t) \Phi_\mathbf{R}(\mathbf{r}_\alpha,t) ]d\mathbf{r}
    + O(M_n^{-1}).
\end{eqnarray}

The TDPES can be further decomposed into gauge dependent and gauge independent components. 
Specifically, the gauge dependent term reads
\begin{eqnarray}\label{TDPES_gd}
\epsilon_{gd}(\mathbf{R},t) =   \frac{\int {\mathcal{D}}_\mathbf{r} [  |\psi_\alpha(\mathbf{R},t)|^2 A_\alpha   ]d\mathbf{r}}{\int {\mathcal{D}}_\mathbf{r}[|\psi_\alpha(\mathbf{R},t)|^2]d\mathbf{r}} \nonumber\\
   +  \int {\mathcal{D}}_\mathbf{r}[\Phi^*_{\mathbf{R}}(\mathbf{r}_\alpha,t) (-i d_t)  \Phi_{\mathbf{R}}(\mathbf{r}_\alpha,t)] d\mathbf{r},
\end{eqnarray}
while the gauge independent part is
\begin{eqnarray}\label{TDPES_gi}
 \epsilon_{gi}(\mathbf{R},t) =   \frac{\int {\mathcal{D}}_\mathbf{r} [  |\psi_\alpha(\mathbf{R},t)|^2 (W_\alpha + K_\alpha)   ]d\mathbf{r}}{\int {\mathcal{D}}_\mathbf{r}[|\psi_\alpha(\mathbf{R},t)|^2]d\mathbf{r}} + O(M_n^{-1}) \nonumber\\
                      = \frac{\int {\mathcal{D}}_\mathbf{r} [  |\psi_\alpha(\mathbf{R},t)|^2 (W_\alpha + K^\Re_\alpha)   ]d\mathbf{r}}{\int {\mathcal{D}}_\mathbf{r}[|\psi_\alpha(\mathbf{R},t)|^2]d\mathbf{r}}  + O(M_n^{-1}), \quad \;
\end{eqnarray}
where in the last equality of (\ref{TDPES_gi}) we made use of the fact  that the imaginary part of $K_\alpha$ vanishes along the integration of the electronic degrees of freedom.

\bibliography{NAMD_CW_BIB}

\begin{thebibliography}{47}%
\makeatletter
\providecommand \@ifxundefined [1]{%
 \@ifx{#1\undefined}
}%
\providecommand \@ifnum [1]{%
 \ifnum #1\expandafter \@firstoftwo
 \else \expandafter \@secondoftwo
 \fi
}%
\providecommand \@ifx [1]{%
 \ifx #1\expandafter \@firstoftwo
 \else \expandafter \@secondoftwo
 \fi
}%
\providecommand \natexlab [1]{#1}%
\providecommand \enquote  [1]{``#1''}%
\providecommand \bibnamefont  [1]{#1}%
\providecommand \bibfnamefont [1]{#1}%
\providecommand \citenamefont [1]{#1}%
\providecommand \href@noop [0]{\@secondoftwo}%
\providecommand \href [0]{\begingroup \@sanitize@url \@href}%
\providecommand \@href[1]{\@@startlink{#1}\@@href}%
\providecommand \@@href[1]{\endgroup#1\@@endlink}%
\providecommand \@sanitize@url [0]{\catcode `\\12\catcode `\$12\catcode
  `\&12\catcode `\#12\catcode `\^12\catcode `\_12\catcode `\%12\relax}%
\providecommand \@@startlink[1]{}%
\providecommand \@@endlink[0]{}%
\providecommand \url  [0]{\begingroup\@sanitize@url \@url }%
\providecommand \@url [1]{\endgroup\@href {#1}{\urlprefix }}%
\providecommand \urlprefix  [0]{URL }%
\providecommand \Eprint [0]{\href }%
\providecommand \doibase [0]{http://dx.doi.org/}%
\providecommand \selectlanguage [0]{\@gobble}%
\providecommand \bibinfo  [0]{\@secondoftwo}%
\providecommand \bibfield  [0]{\@secondoftwo}%
\providecommand \translation [1]{[#1]}%
\providecommand \BibitemOpen [0]{}%
\providecommand \bibitemStop [0]{}%
\providecommand \bibitemNoStop [0]{.\EOS\space}%
\providecommand \EOS [0]{\spacefactor3000\relax}%
\providecommand \BibitemShut  [1]{\csname bibitem#1\endcsname}%
\let\auto@bib@innerbib\@empty
\bibitem [{\citenamefont {Clark}\ \emph {et~al.}(2012)\citenamefont {Clark},
  \citenamefont {Nelson}, \citenamefont {Tretiak}, \citenamefont {Cirmi},\ and\
  \citenamefont {Lanzani}}]{exp1}%
  \BibitemOpen
  \bibfield  {author} {\bibinfo {author} {\bibfnamefont {J.}~\bibnamefont
  {Clark}}, \bibinfo {author} {\bibfnamefont {T.}~\bibnamefont {Nelson}},
  \bibinfo {author} {\bibfnamefont {S.}~\bibnamefont {Tretiak}}, \bibinfo
  {author} {\bibfnamefont {G.}~\bibnamefont {Cirmi}}, \ and\ \bibinfo {author}
  {\bibfnamefont {G.}~\bibnamefont {Lanzani}},\ }\href@noop {} {\bibfield
  {journal} {\bibinfo  {journal} {Nature Physics}\ }\textbf {\bibinfo {volume}
  {8}},\ \bibinfo {pages} {225} (\bibinfo {year} {2012})}\BibitemShut {NoStop}%
\bibitem [{\citenamefont {Zhou}\ \emph {et~al.}(2012)\citenamefont {Zhou},
  \citenamefont {Ranitovic}, \citenamefont {Hogle}, \citenamefont {Eland},
  \citenamefont {Kapteyn},\ and\ \citenamefont {Murnane}}]{exp2}%
  \BibitemOpen
  \bibfield  {author} {\bibinfo {author} {\bibfnamefont {X.}~\bibnamefont
  {Zhou}}, \bibinfo {author} {\bibfnamefont {P.}~\bibnamefont {Ranitovic}},
  \bibinfo {author} {\bibfnamefont {C.}~\bibnamefont {Hogle}}, \bibinfo
  {author} {\bibfnamefont {J.}~\bibnamefont {Eland}}, \bibinfo {author}
  {\bibfnamefont {H.}~\bibnamefont {Kapteyn}}, \ and\ \bibinfo {author}
  {\bibfnamefont {M.}~\bibnamefont {Murnane}},\ }\href@noop {} {\bibfield
  {journal} {\bibinfo  {journal} {Nature Physics}\ }\textbf {\bibinfo {volume}
  {8}},\ \bibinfo {pages} {232} (\bibinfo {year} {2012})}\BibitemShut {NoStop}%
\bibitem [{\citenamefont {Leone}\ \emph {et~al.}(2014)\citenamefont {Leone},
  \citenamefont {McCurdy}, \citenamefont {Burgd{\"o}rfer}, \citenamefont
  {Cederbaum}, \citenamefont {Chang}, \citenamefont {Dudovich}, \citenamefont
  {Feist}, \citenamefont {Greene}, \citenamefont {Ivanov}, \citenamefont
  {Kienberger} \emph {et~al.}}]{exp3}%
  \BibitemOpen
  \bibfield  {author} {\bibinfo {author} {\bibfnamefont {S.~R.}\ \bibnamefont
  {Leone}}, \bibinfo {author} {\bibfnamefont {C.~W.}\ \bibnamefont {McCurdy}},
  \bibinfo {author} {\bibfnamefont {J.}~\bibnamefont {Burgd{\"o}rfer}},
  \bibinfo {author} {\bibfnamefont {L.~S.}\ \bibnamefont {Cederbaum}}, \bibinfo
  {author} {\bibfnamefont {Z.}~\bibnamefont {Chang}}, \bibinfo {author}
  {\bibfnamefont {N.}~\bibnamefont {Dudovich}}, \bibinfo {author}
  {\bibfnamefont {J.}~\bibnamefont {Feist}}, \bibinfo {author} {\bibfnamefont
  {C.~H.}\ \bibnamefont {Greene}}, \bibinfo {author} {\bibfnamefont
  {M.}~\bibnamefont {Ivanov}}, \bibinfo {author} {\bibfnamefont
  {R.}~\bibnamefont {Kienberger}},  \emph {et~al.},\ }\href@noop {} {\bibfield
  {journal} {\bibinfo  {journal} {Nature Photonics}\ }\textbf {\bibinfo
  {volume} {8}},\ \bibinfo {pages} {162} (\bibinfo {year} {2014})}\BibitemShut
  {NoStop}%
\bibitem [{\citenamefont {L{\'e}pine}\ \emph {et~al.}(2014)\citenamefont
  {L{\'e}pine}, \citenamefont {Ivanov},\ and\ \citenamefont {Vrakking}}]{exp4}%
  \BibitemOpen
  \bibfield  {author} {\bibinfo {author} {\bibfnamefont {F.}~\bibnamefont
  {L{\'e}pine}}, \bibinfo {author} {\bibfnamefont {M.~Y.}\ \bibnamefont
  {Ivanov}}, \ and\ \bibinfo {author} {\bibfnamefont {M.~J.}\ \bibnamefont
  {Vrakking}},\ }\href@noop {} {\bibfield  {journal} {\bibinfo  {journal}
  {Nature Photonics}\ }\textbf {\bibinfo {volume} {8}},\ \bibinfo {pages} {195}
  (\bibinfo {year} {2014})}\BibitemShut {NoStop}%
\bibitem [{\citenamefont {Boge}\ \emph {et~al.}(2013)\citenamefont {Boge},
  \citenamefont {Cirelli}, \citenamefont {Landsman}, \citenamefont {Heuser},
  \citenamefont {Ludwig}, \citenamefont {Maurer}, \citenamefont {Weger},
  \citenamefont {Gallmann},\ and\ \citenamefont {Keller}}]{exp5}%
  \BibitemOpen
  \bibfield  {author} {\bibinfo {author} {\bibfnamefont {R.}~\bibnamefont
  {Boge}}, \bibinfo {author} {\bibfnamefont {C.}~\bibnamefont {Cirelli}},
  \bibinfo {author} {\bibfnamefont {A.}~\bibnamefont {Landsman}}, \bibinfo
  {author} {\bibfnamefont {S.}~\bibnamefont {Heuser}}, \bibinfo {author}
  {\bibfnamefont {A.}~\bibnamefont {Ludwig}}, \bibinfo {author} {\bibfnamefont
  {J.}~\bibnamefont {Maurer}}, \bibinfo {author} {\bibfnamefont
  {M.}~\bibnamefont {Weger}}, \bibinfo {author} {\bibfnamefont
  {L.}~\bibnamefont {Gallmann}}, \ and\ \bibinfo {author} {\bibfnamefont
  {U.}~\bibnamefont {Keller}},\ }\href@noop {} {\bibfield  {journal} {\bibinfo
  {journal} {Physical review letters}\ }\textbf {\bibinfo {volume} {111}},\
  \bibinfo {pages} {103003} (\bibinfo {year} {2013})}\BibitemShut {NoStop}%
\bibitem [{\citenamefont {Corrales}\ \emph {et~al.}(2014)\citenamefont
  {Corrales}, \citenamefont {Gonz{\'a}lez-V{\'a}zquez}, \citenamefont
  {Balerdi}, \citenamefont {Sol{\'a}}, \citenamefont {de~Nalda},\ and\
  \citenamefont {Ba{\~n}ares}}]{exp6}%
  \BibitemOpen
  \bibfield  {author} {\bibinfo {author} {\bibfnamefont {M.}~\bibnamefont
  {Corrales}}, \bibinfo {author} {\bibfnamefont {J.}~\bibnamefont
  {Gonz{\'a}lez-V{\'a}zquez}}, \bibinfo {author} {\bibfnamefont
  {G.}~\bibnamefont {Balerdi}}, \bibinfo {author} {\bibfnamefont
  {I.}~\bibnamefont {Sol{\'a}}}, \bibinfo {author} {\bibfnamefont
  {R.}~\bibnamefont {de~Nalda}}, \ and\ \bibinfo {author} {\bibfnamefont
  {L.}~\bibnamefont {Ba{\~n}ares}},\ }\href@noop {} {\bibfield  {journal}
  {\bibinfo  {journal} {Nature chemistry}\ }\textbf {\bibinfo {volume} {6}},\
  \bibinfo {pages} {785} (\bibinfo {year} {2014})}\BibitemShut {NoStop}%
\bibitem [{\citenamefont {Tavernelli}(2015)}]{rev1}%
  \BibitemOpen
  \bibfield  {author} {\bibinfo {author} {\bibfnamefont {I.}~\bibnamefont
  {Tavernelli}},\ }\href@noop {} {\bibfield  {journal} {\bibinfo  {journal}
  {Accounts of chemical research}\ }\textbf {\bibinfo {volume} {48}},\ \bibinfo
  {pages} {792} (\bibinfo {year} {2015})}\BibitemShut {NoStop}%
\bibitem [{\citenamefont {Persico}\ and\ \citenamefont
  {Granucci}(2014{\natexlab{a}})}]{rev2}%
  \BibitemOpen
  \bibfield  {author} {\bibinfo {author} {\bibfnamefont {M.}~\bibnamefont
  {Persico}}\ and\ \bibinfo {author} {\bibfnamefont {G.}~\bibnamefont
  {Granucci}},\ }\href@noop {} {\bibfield  {journal} {\bibinfo  {journal}
  {Theoretical Chemistry Accounts}\ }\textbf {\bibinfo {volume} {133}},\
  \bibinfo {pages} {1} (\bibinfo {year} {2014}{\natexlab{a}})}\BibitemShut
  {NoStop}%
\bibitem [{\citenamefont {Lasorne}\ \emph {et~al.}(2014)\citenamefont
  {Lasorne}, \citenamefont {Worth},\ and\ \citenamefont {Robb}}]{rev3}%
  \BibitemOpen
  \bibfield  {author} {\bibinfo {author} {\bibfnamefont {B.}~\bibnamefont
  {Lasorne}}, \bibinfo {author} {\bibfnamefont {G.~A.}\ \bibnamefont {Worth}},
  \ and\ \bibinfo {author} {\bibfnamefont {M.~A.}\ \bibnamefont {Robb}},\ }in\
  \href@noop {} {\emph {\bibinfo {booktitle} {Molecular Quantum Dynamics}}}\
  (\bibinfo  {publisher} {Springer},\ \bibinfo {year} {2014})\ pp.\ \bibinfo
  {pages} {181--211}\BibitemShut {NoStop}%
\bibitem [{\citenamefont {Domcke}\ and\ \citenamefont {Yarkony}(2012)}]{rev4}%
  \BibitemOpen
  \bibfield  {author} {\bibinfo {author} {\bibfnamefont {W.}~\bibnamefont
  {Domcke}}\ and\ \bibinfo {author} {\bibfnamefont {D.~R.}\ \bibnamefont
  {Yarkony}},\ }\href@noop {} {\bibfield  {journal} {\bibinfo  {journal}
  {Annual review of physical chemistry}\ }\textbf {\bibinfo {volume} {63}},\
  \bibinfo {pages} {325} (\bibinfo {year} {2012})}\BibitemShut {NoStop}%
\bibitem [{\citenamefont {Tully}(2012)}]{rev5}%
  \BibitemOpen
  \bibfield  {author} {\bibinfo {author} {\bibfnamefont {J.~C.}\ \bibnamefont
  {Tully}},\ }\href@noop {} {\bibfield  {journal} {\bibinfo  {journal} {The
  Journal of chemical physics}\ }\textbf {\bibinfo {volume} {137}},\ \bibinfo
  {pages} {22A301} (\bibinfo {year} {2012})}\BibitemShut {NoStop}%
\bibitem [{\citenamefont {Truhlar}(1995)}]{truhlar}%
  \BibitemOpen
  \bibfield  {author} {\bibinfo {author} {\bibfnamefont {D.}~\bibnamefont
  {Truhlar}},\ }in\ \href {http://dx.doi.org/10.1007/978-94-015-8539-2_10}
  {\emph {\bibinfo {booktitle} {The Reaction Path in Chemistry: Current
  Approaches and Perspectives}}},\ \bibinfo {series} {Understanding Chemical
  Reactivity}, Vol.~\bibinfo {volume} {16},\ \bibinfo {editor} {edited by\
  \bibinfo {editor} {\bibfnamefont {D.}~\bibnamefont {Heidrich}}}\ (\bibinfo
  {publisher} {Springer Netherlands},\ \bibinfo {year} {1995})\ pp.\ \bibinfo
  {pages} {229--255}\BibitemShut {NoStop}%
\bibitem [{\citenamefont {Curchod}\ \emph {et~al.}(2013)\citenamefont
  {Curchod}, \citenamefont {Rothlisberger},\ and\ \citenamefont
  {Tavernelli}}]{Tavernelli2}%
  \BibitemOpen
  \bibfield  {author} {\bibinfo {author} {\bibfnamefont {B.~F.~E.}\
  \bibnamefont {Curchod}}, \bibinfo {author} {\bibfnamefont {U.}~\bibnamefont
  {Rothlisberger}}, \ and\ \bibinfo {author} {\bibfnamefont {I.}~\bibnamefont
  {Tavernelli}},\ }\href {http://dx.doi.org/10.1002/cphc.201200941} {\bibfield
  {journal} {\bibinfo  {journal} {ChemPhysChem}\ }\textbf {\bibinfo {volume}
  {14}},\ \bibinfo {pages} {1314} (\bibinfo {year} {2013})}\BibitemShut
  {NoStop}%
\bibitem [{\citenamefont {Tavernelli}(2013)}]{Tavernelli3}%
  \BibitemOpen
  \bibfield  {author} {\bibinfo {author} {\bibfnamefont {I.}~\bibnamefont
  {Tavernelli}},\ }\href {\doibase 10.1103/PhysRevA.87.042501} {\bibfield
  {journal} {\bibinfo  {journal} {Phys. Rev. A}\ }\textbf {\bibinfo {volume}
  {87}},\ \bibinfo {pages} {042501} (\bibinfo {year} {2013})}\BibitemShut
  {NoStop}%
\bibitem [{\citenamefont {Tully}(1990)}]{SH}%
  \BibitemOpen
  \bibfield  {author} {\bibinfo {author} {\bibfnamefont {J.~C.}\ \bibnamefont
  {Tully}},\ }\href
  {http://scitation.aip.org/content/aip/journal/jcp/93/2/10.1063/1.459170}
  {\bibfield  {journal} {\bibinfo  {journal} {J. Chem. Phys.}\ }\textbf
  {\bibinfo {volume} {93}},\ \bibinfo {pages} {1061} (\bibinfo {year}
  {1990})}\BibitemShut {NoStop}%
\bibitem [{\citenamefont {C.~Tully}(1998)}]{tully1998mixed}%
  \BibitemOpen
  \bibfield  {author} {\bibinfo {author} {\bibfnamefont {J.}~\bibnamefont
  {C.~Tully}},\ }\href {http://dx.doi.org/10.1039/A801824C} {\bibfield
  {journal} {\bibinfo  {journal} {Farad. Discuss.}\ }\textbf {\bibinfo {volume}
  {110}},\ \bibinfo {pages} {407} (\bibinfo {year} {1998})}\BibitemShut
  {NoStop}%
\bibitem [{\citenamefont {Kapral}\ and\ \citenamefont
  {Ciccotti}(1999)}]{kapral1999mixed}%
  \BibitemOpen
  \bibfield  {author} {\bibinfo {author} {\bibfnamefont {R.}~\bibnamefont
  {Kapral}}\ and\ \bibinfo {author} {\bibfnamefont {G.}~\bibnamefont
  {Ciccotti}},\ }\href@noop {} {\bibfield  {journal} {\bibinfo  {journal} {The
  Journal of chemical physics}\ }\textbf {\bibinfo {volume} {110}},\ \bibinfo
  {pages} {8919} (\bibinfo {year} {1999})}\BibitemShut {NoStop}%
\bibitem [{\citenamefont {Persico}\ and\ \citenamefont
  {Granucci}(2014{\natexlab{b}})}]{persico2014overview}%
  \BibitemOpen
  \bibfield  {author} {\bibinfo {author} {\bibfnamefont {M.}~\bibnamefont
  {Persico}}\ and\ \bibinfo {author} {\bibfnamefont {G.}~\bibnamefont
  {Granucci}},\ }\href@noop {} {\bibfield  {journal} {\bibinfo  {journal}
  {Theoretical Chemistry Accounts}\ }\textbf {\bibinfo {volume} {133}},\
  \bibinfo {pages} {1} (\bibinfo {year} {2014}{\natexlab{b}})}\BibitemShut
  {NoStop}%
\bibitem [{\citenamefont {Sawada}\ and\ \citenamefont
  {Metiu}(1986)}]{SawadaMetiu}%
  \BibitemOpen
  \bibfield  {author} {\bibinfo {author} {\bibfnamefont {S.}~\bibnamefont
  {Sawada}}\ and\ \bibinfo {author} {\bibfnamefont {H.}~\bibnamefont {Metiu}},\
  }\href
  {http://scitation.aip.org/content/aip/journal/jcp/84/1/10.1063/1.450175}
  {\bibfield  {journal} {\bibinfo  {journal} {J. Chem. Phys.}\ }\textbf
  {\bibinfo {volume} {84}},\ \bibinfo {pages} {227} (\bibinfo {year}
  {1986})}\BibitemShut {NoStop}%
\bibitem [{\citenamefont {Martinez}\ \emph {et~al.}(1996)\citenamefont
  {Martinez}, \citenamefont {Ben-Nun},\ and\ \citenamefont
  {Levine}}]{Martinez}%
  \BibitemOpen
  \bibfield  {author} {\bibinfo {author} {\bibfnamefont {T.~J.}\ \bibnamefont
  {Martinez}}, \bibinfo {author} {\bibfnamefont {M.}~\bibnamefont {Ben-Nun}}, \
  and\ \bibinfo {author} {\bibfnamefont {R.~D.}\ \bibnamefont {Levine}},\
  }\href {http://pubs.acs.org/doi/abs/10.1021/jp953105a} {\bibfield  {journal}
  {\bibinfo  {journal} {J. Phys. Chem.}\ }\textbf {\bibinfo {volume} {100}},\
  \bibinfo {pages} {7884} (\bibinfo {year} {1996})}\BibitemShut {NoStop}%
\bibitem [{\citenamefont {Zheng}\ \emph {et~al.}(2014)\citenamefont {Zheng},
  \citenamefont {Xu}, \citenamefont {Meana-Paneda},\ and\ \citenamefont
  {Truhlar}}]{truhlar_ants}%
  \BibitemOpen
  \bibfield  {author} {\bibinfo {author} {\bibfnamefont {J.}~\bibnamefont
  {Zheng}}, \bibinfo {author} {\bibfnamefont {X.}~\bibnamefont {Xu}}, \bibinfo
  {author} {\bibfnamefont {R.}~\bibnamefont {Meana-Paneda}}, \ and\ \bibinfo
  {author} {\bibfnamefont {D.~G.}\ \bibnamefont {Truhlar}},\ }\href {\doibase
  10.1039/C3SC53290A} {\bibfield  {journal} {\bibinfo  {journal} {Chem. Sci.}\
  }\textbf {\bibinfo {volume} {5}},\ \bibinfo {pages} {2091} (\bibinfo {year}
  {2014})}\BibitemShut {NoStop}%
\bibitem [{\citenamefont {Burghardt}\ \emph {et~al.}(1999)\citenamefont
  {Burghardt}, \citenamefont {Meyer},\ and\ \citenamefont
  {Cederbaum}}]{BurghardtCederbaum}%
  \BibitemOpen
  \bibfield  {author} {\bibinfo {author} {\bibfnamefont {I.}~\bibnamefont
  {Burghardt}}, \bibinfo {author} {\bibfnamefont {H.-D.}\ \bibnamefont
  {Meyer}}, \ and\ \bibinfo {author} {\bibfnamefont {L.~S.}\ \bibnamefont
  {Cederbaum}},\ }\href
  {http://scitation.aip.org/content/aip/journal/jcp/111/7/10.1063/1.479574}
  {\bibfield  {journal} {\bibinfo  {journal} {J. Chem. Phys.}\ }\textbf
  {\bibinfo {volume} {111}},\ \bibinfo {pages} {2927} (\bibinfo {year}
  {1999})}\BibitemShut {NoStop}%
\bibitem [{\citenamefont {Beck}\ \emph {et~al.}(2000)\citenamefont {Beck},
  \citenamefont {Jäckle}, \citenamefont {Worth},\ and\ \citenamefont
  {Meyer}}]{MCTDH1}%
  \BibitemOpen
  \bibfield  {author} {\bibinfo {author} {\bibfnamefont {M.}~\bibnamefont
  {Beck}}, \bibinfo {author} {\bibfnamefont {A.}~\bibnamefont {Jäckle}},
  \bibinfo {author} {\bibfnamefont {G.}~\bibnamefont {Worth}}, \ and\ \bibinfo
  {author} {\bibfnamefont {H.-D.}\ \bibnamefont {Meyer}},\ }\href
  {http://www.sciencedirect.com/science/article/pii/S0370157399000472}
  {\bibfield  {journal} {\bibinfo  {journal} {Phys. Rep.}\ }\textbf {\bibinfo
  {volume} {324}},\ \bibinfo {pages} {1 } (\bibinfo {year} {2000})}\BibitemShut
  {NoStop}%
\bibitem [{\citenamefont {Meyer}\ and\ \citenamefont {Worth}(2003)}]{MCTDH2}%
  \BibitemOpen
  \bibfield  {author} {\bibinfo {author} {\bibfnamefont {H.-D.}\ \bibnamefont
  {Meyer}}\ and\ \bibinfo {author} {\bibfnamefont {G.~A.}\ \bibnamefont
  {Worth}},\ }\href {\doibase 10.1007/s00214-003-0439-1} {\bibfield  {journal}
  {\bibinfo  {journal} {Theor. Chem. Acc.}\ }\textbf {\bibinfo {volume}
  {109}},\ \bibinfo {pages} {251} (\bibinfo {year} {2003})}\BibitemShut
  {NoStop}%
\bibitem [{\citenamefont {Abedi}\ \emph {et~al.}(2010)\citenamefont {Abedi},
  \citenamefont {Maitra},\ and\ \citenamefont {Gross}}]{Ali1}%
  \BibitemOpen
  \bibfield  {author} {\bibinfo {author} {\bibfnamefont {A.}~\bibnamefont
  {Abedi}}, \bibinfo {author} {\bibfnamefont {N.~T.}\ \bibnamefont {Maitra}}, \
  and\ \bibinfo {author} {\bibfnamefont {E.~K.~U.}\ \bibnamefont {Gross}},\
  }\href {http://link.aps.org/doi/10.1103/PhysRevLett.105.123002} {\bibfield
  {journal} {\bibinfo  {journal} {Phys. Rev. Lett.}\ }\textbf {\bibinfo
  {volume} {105}},\ \bibinfo {pages} {123002} (\bibinfo {year}
  {2010})}\BibitemShut {NoStop}%
\bibitem [{\citenamefont {Albareda}\ \emph {et~al.}(2014)\citenamefont
  {Albareda}, \citenamefont {Appel}, \citenamefont {Franco}, \citenamefont
  {Abedi},\ and\ \citenamefont {Rubio}}]{Guil1}%
  \BibitemOpen
  \bibfield  {author} {\bibinfo {author} {\bibfnamefont {G.}~\bibnamefont
  {Albareda}}, \bibinfo {author} {\bibfnamefont {H.}~\bibnamefont {Appel}},
  \bibinfo {author} {\bibfnamefont {I.}~\bibnamefont {Franco}}, \bibinfo
  {author} {\bibfnamefont {A.}~\bibnamefont {Abedi}}, \ and\ \bibinfo {author}
  {\bibfnamefont {A.}~\bibnamefont {Rubio}},\ }\href {\doibase
  10.1103/PhysRevLett.113.083003} {\bibfield  {journal} {\bibinfo  {journal}
  {Phys. Rev. Lett.}\ }\textbf {\bibinfo {volume} {113}},\ \bibinfo {pages}
  {083003} (\bibinfo {year} {2014})}\BibitemShut {NoStop}%
\bibitem [{\citenamefont {Alonso}\ \emph {et~al.}(2008)\citenamefont {Alonso},
  \citenamefont {Andrade}, \citenamefont {Echenique}, \citenamefont {Falceto},
  \citenamefont {Prada-Gracia},\ and\ \citenamefont
  {Rubio}}]{alonso2008Ehrenfest}%
  \BibitemOpen
  \bibfield  {author} {\bibinfo {author} {\bibfnamefont {J.~L.}\ \bibnamefont
  {Alonso}}, \bibinfo {author} {\bibfnamefont {X.}~\bibnamefont {Andrade}},
  \bibinfo {author} {\bibfnamefont {P.}~\bibnamefont {Echenique}}, \bibinfo
  {author} {\bibfnamefont {F.}~\bibnamefont {Falceto}}, \bibinfo {author}
  {\bibfnamefont {D.}~\bibnamefont {Prada-Gracia}}, \ and\ \bibinfo {author}
  {\bibfnamefont {A.}~\bibnamefont {Rubio}},\ }\href@noop {} {\bibfield
  {journal} {\bibinfo  {journal} {Physical review letters}\ }\textbf {\bibinfo
  {volume} {101}},\ \bibinfo {pages} {096403} (\bibinfo {year}
  {2008})}\BibitemShut {NoStop}%
\bibitem [{\citenamefont {McEniry}\ \emph {et~al.}(2010)\citenamefont
  {McEniry}, \citenamefont {Wang}, \citenamefont {Dundas}, \citenamefont
  {Todorov}, \citenamefont {Stella}, \citenamefont {Miranda}, \citenamefont
  {Fisher}, \citenamefont {Horsfield}, \citenamefont {Race}, \citenamefont
  {Mason} \emph {et~al.}}]{mceniry2010Ehrenfest}%
  \BibitemOpen
  \bibfield  {author} {\bibinfo {author} {\bibfnamefont {E.}~\bibnamefont
  {McEniry}}, \bibinfo {author} {\bibfnamefont {Y.}~\bibnamefont {Wang}},
  \bibinfo {author} {\bibfnamefont {D.}~\bibnamefont {Dundas}}, \bibinfo
  {author} {\bibfnamefont {T.}~\bibnamefont {Todorov}}, \bibinfo {author}
  {\bibfnamefont {L.}~\bibnamefont {Stella}}, \bibinfo {author} {\bibfnamefont
  {R.}~\bibnamefont {Miranda}}, \bibinfo {author} {\bibfnamefont
  {A.}~\bibnamefont {Fisher}}, \bibinfo {author} {\bibfnamefont
  {A.}~\bibnamefont {Horsfield}}, \bibinfo {author} {\bibfnamefont
  {C.}~\bibnamefont {Race}}, \bibinfo {author} {\bibfnamefont {D.}~\bibnamefont
  {Mason}},  \emph {et~al.},\ }\href@noop {} {\bibfield  {journal} {\bibinfo
  {journal} {The European Physical Journal B}\ }\textbf {\bibinfo {volume}
  {77}},\ \bibinfo {pages} {305} (\bibinfo {year} {2010})}\BibitemShut
  {NoStop}%
\bibitem [{\citenamefont {Horsfield}\ \emph {et~al.}(2004)\citenamefont
  {Horsfield}, \citenamefont {Bowler}, \citenamefont {Fisher}, \citenamefont
  {Todorov},\ and\ \citenamefont {S{\'a}nchez}}]{horsfield2004Ehrenfest}%
  \BibitemOpen
  \bibfield  {author} {\bibinfo {author} {\bibfnamefont {A.~P.}\ \bibnamefont
  {Horsfield}}, \bibinfo {author} {\bibfnamefont {D.}~\bibnamefont {Bowler}},
  \bibinfo {author} {\bibfnamefont {A.}~\bibnamefont {Fisher}}, \bibinfo
  {author} {\bibfnamefont {T.~N.}\ \bibnamefont {Todorov}}, \ and\ \bibinfo
  {author} {\bibfnamefont {C.~G.}\ \bibnamefont {S{\'a}nchez}},\ }\href@noop {}
  {\bibfield  {journal} {\bibinfo  {journal} {Journal of Physics: Condensed
  Matter}\ }\textbf {\bibinfo {volume} {16}},\ \bibinfo {pages} {8251}
  (\bibinfo {year} {2004})}\BibitemShut {NoStop}%
\bibitem [{Note1()}]{Note1}%
  \BibitemOpen
  \bibinfo {note} {The electronic support of the full probability density
  $|\Psi (\protect \mathbf {r},\protect \mathbf {R},t)|^2$ can be defined as:
  $\protect \textit {supp}(|\Psi (\protect \mathbf {r},\protect \mathbf
  {R},t)|^2) = \protect \{ \protect \mathfrak {r} \in \protect \mathbf {r}
  \mskip \thickmuskip \protect \mathbf {|} \mskip \thickmuskip |\Psi (\protect
  \mathbf {r},\protect \mathbf {R},t)|^2 > \lambda \protect \}$, where $\lambda
  $ denotes a given lower bound of the probability density.}\BibitemShut
  {Stop}%
\bibitem [{Note2()}]{Note2}%
  \BibitemOpen
  \bibinfo {note} {{ This can be easily done from $\Psi (\protect \mathbf {r},
  \protect \mathbf {R} ,t) = {\protect \mathcal {D}}_\protect \mathbf {r} [\psi
  _\alpha ]$, where the transformation ${\protect \mathcal {D}}$ has been
  defined in Eq.~(\ref {transform}).}}\BibitemShut {Stop}%
\bibitem [{\citenamefont {Albareda}\ \emph {et~al.}(2015)\citenamefont
  {Albareda}, \citenamefont {Bofill}, \citenamefont {Tavernelli}, \citenamefont
  {Huarte-Larrañaga}, \citenamefont {Illas},\ and\ \citenamefont
  {Rubio}}]{JPCL}%
  \BibitemOpen
  \bibfield  {author} {\bibinfo {author} {\bibfnamefont {G.}~\bibnamefont
  {Albareda}}, \bibinfo {author} {\bibfnamefont {J.~M.}\ \bibnamefont
  {Bofill}}, \bibinfo {author} {\bibfnamefont {I.}~\bibnamefont {Tavernelli}},
  \bibinfo {author} {\bibfnamefont {F.}~\bibnamefont {Huarte-Larrañaga}},
  \bibinfo {author} {\bibfnamefont {F.}~\bibnamefont {Illas}}, \ and\ \bibinfo
  {author} {\bibfnamefont {A.}~\bibnamefont {Rubio}},\ }\href@noop {}
  {\bibfield  {journal} {\bibinfo  {journal} {The Journal of Physical Chemistry
  Letters}\ }\textbf {\bibinfo {volume} {6}},\ \bibinfo {pages} {1529}
  (\bibinfo {year} {2015})}\BibitemShut {NoStop}%
\bibitem [{\citenamefont {Albareda}\ \emph {et~al.}(2009)\citenamefont
  {Albareda}, \citenamefont {Su\~n\'e},\ and\ \citenamefont
  {Oriols}}]{PRBGuille}%
  \BibitemOpen
  \bibfield  {author} {\bibinfo {author} {\bibfnamefont {G.}~\bibnamefont
  {Albareda}}, \bibinfo {author} {\bibfnamefont {J.}~\bibnamefont {Su\~n\'e}},
  \ and\ \bibinfo {author} {\bibfnamefont {X.}~\bibnamefont {Oriols}},\ }\href
  {http://link.aps.org/doi/10.1103/PhysRevB.79.075315} {\bibfield  {journal}
  {\bibinfo  {journal} {Phys. Rev. B}\ }\textbf {\bibinfo {volume} {79}},\
  \bibinfo {pages} {075315} (\bibinfo {year} {2009})}\BibitemShut {NoStop}%
\bibitem [{\citenamefont {Albareda}\ \emph {et~al.}(2013)\citenamefont
  {Albareda}, \citenamefont {Marian}, \citenamefont {Benali}, \citenamefont
  {Yaro}, \citenamefont {Zanghì},\ and\ \citenamefont {Oriols}}]{JCE}%
  \BibitemOpen
  \bibfield  {author} {\bibinfo {author} {\bibfnamefont {G.}~\bibnamefont
  {Albareda}}, \bibinfo {author} {\bibfnamefont {D.}~\bibnamefont {Marian}},
  \bibinfo {author} {\bibfnamefont {A.}~\bibnamefont {Benali}}, \bibinfo
  {author} {\bibfnamefont {S.}~\bibnamefont {Yaro}}, \bibinfo {author}
  {\bibfnamefont {N.}~\bibnamefont {Zanghì}}, \ and\ \bibinfo {author}
  {\bibfnamefont {X.}~\bibnamefont {Oriols}},\ }\href {\doibase
  10.1007/s10825-013-0484-5} {\bibfield  {journal} {\bibinfo  {journal} {J.
  Comp. Electr.}\ }\textbf {\bibinfo {volume} {12}},\ \bibinfo {pages} {405}
  (\bibinfo {year} {2013})}\BibitemShut {NoStop}%
\bibitem [{\citenamefont {Oriols}(2007)}]{PRLOriols}%
  \BibitemOpen
  \bibfield  {author} {\bibinfo {author} {\bibfnamefont {X.}~\bibnamefont
  {Oriols}},\ }\href {http://link.aps.org/doi/10.1103/PhysRevLett.98.066803}
  {\bibfield  {journal} {\bibinfo  {journal} {Phys. Rev. Lett.}\ }\textbf
  {\bibinfo {volume} {98}},\ \bibinfo {pages} {066803} (\bibinfo {year}
  {2007})}\BibitemShut {NoStop}%
\bibitem [{\citenamefont {Albareda}\ \emph {et~al.}(2016)\citenamefont
  {Albareda}, \citenamefont {Marian}, \citenamefont {Benali}, \citenamefont
  {Alarc{\'o}n}, \citenamefont {Moises},\ and\ \citenamefont
  {Oriols}}]{albareda2016bitlles}%
  \BibitemOpen
  \bibfield  {author} {\bibinfo {author} {\bibfnamefont {G.}~\bibnamefont
  {Albareda}}, \bibinfo {author} {\bibfnamefont {D.}~\bibnamefont {Marian}},
  \bibinfo {author} {\bibfnamefont {A.}~\bibnamefont {Benali}}, \bibinfo
  {author} {\bibfnamefont {A.}~\bibnamefont {Alarc{\'o}n}}, \bibinfo {author}
  {\bibfnamefont {S.}~\bibnamefont {Moises}}, \ and\ \bibinfo {author}
  {\bibfnamefont {X.}~\bibnamefont {Oriols}},\ }\href@noop {} {\bibfield
  {journal} {\bibinfo  {journal} {arXiv preprint arXiv:1609.06534}\ } (\bibinfo
  {year} {2016})}\BibitemShut {NoStop}%
\bibitem [{\citenamefont {Benseny}\ \emph {et~al.}(2014)\citenamefont
  {Benseny}, \citenamefont {Albareda}, \citenamefont {Sanz}, \citenamefont
  {Mompart},\ and\ \citenamefont {Oriols}}]{AppBohm}%
  \BibitemOpen
  \bibfield  {author} {\bibinfo {author} {\bibfnamefont {A.}~\bibnamefont
  {Benseny}}, \bibinfo {author} {\bibfnamefont {G.}~\bibnamefont {Albareda}},
  \bibinfo {author} {\bibfnamefont {Ã.~S.}\ \bibnamefont {Sanz}}, \bibinfo
  {author} {\bibfnamefont {J.}~\bibnamefont {Mompart}}, \ and\ \bibinfo
  {author} {\bibfnamefont {X.}~\bibnamefont {Oriols}},\ }\href {\doibase
  10.1140/epjd/e2014-50222-4} {\bibfield  {journal} {\bibinfo  {journal} {The
  European Physical Journal D}\ }\textbf {\bibinfo {volume} {68}},\ \bibinfo
  {eid} {286} (\bibinfo {year} {2014}),\
  10.1140/epjd/e2014-50222-4}\BibitemShut {NoStop}%
\bibitem [{\citenamefont {Oriols}\ and\ \citenamefont
  {Mompart}(2012)}]{BookOriols}%
  \BibitemOpen
  \bibfield  {author} {\bibinfo {author} {\bibfnamefont {X.}~\bibnamefont
  {Oriols}}\ and\ \bibinfo {author} {\bibfnamefont {J.}~\bibnamefont
  {Mompart}},\ }\href {http://books.google.de/books?id=mnqNx66amcIC} {\emph
  {\bibinfo {title} {Applied Bohmian Mechanics: From Nanoscale Systems to
  Cosmology}}}\ (\bibinfo  {publisher} {Pan Stanford},\ \bibinfo {year}
  {2012})\BibitemShut {NoStop}%
\bibitem [{Note3()}]{Note3}%
  \BibitemOpen
  \bibinfo {note} {Notice that the summation over $\alpha $ implies the
  discretization of the electronic subspace into a countable (infinite) number
  of points. While the support of $ \Psi (\protect \mathbf {r},\protect \mathbf
  {R},t)$ consists of an countless (infinite) number of points, we have assumed
  here the discretization of the configuration space on a numerical
  grid.}\BibitemShut {Stop}%
\bibitem [{\citenamefont {Abedi}\ \emph {et~al.}(2013)\citenamefont {Abedi},
  \citenamefont {Agostini}, \citenamefont {Suzuki},\ and\ \citenamefont
  {Gross}}]{Ali2}%
  \BibitemOpen
  \bibfield  {author} {\bibinfo {author} {\bibfnamefont {A.}~\bibnamefont
  {Abedi}}, \bibinfo {author} {\bibfnamefont {F.}~\bibnamefont {Agostini}},
  \bibinfo {author} {\bibfnamefont {Y.}~\bibnamefont {Suzuki}}, \ and\ \bibinfo
  {author} {\bibfnamefont {E.~K.~U.}\ \bibnamefont {Gross}},\ }\href
  {http://link.aps.org/doi/10.1103/PhysRevLett.110.263001} {\bibfield
  {journal} {\bibinfo  {journal} {Phys. Rev. Lett.}\ }\textbf {\bibinfo
  {volume} {110}},\ \bibinfo {pages} {263001} (\bibinfo {year}
  {2013})}\BibitemShut {NoStop}%
\bibitem [{\citenamefont {Booth}\ \emph {et~al.}(2009)\citenamefont {Booth},
  \citenamefont {Thom},\ and\ \citenamefont {Alavi}}]{deadwood}%
  \BibitemOpen
  \bibfield  {author} {\bibinfo {author} {\bibfnamefont {G.~H.}\ \bibnamefont
  {Booth}}, \bibinfo {author} {\bibfnamefont {A.~J.~W.}\ \bibnamefont {Thom}},
  \ and\ \bibinfo {author} {\bibfnamefont {A.}~\bibnamefont {Alavi}},\
  }\href@noop {} {\bibfield  {journal} {\bibinfo  {journal} {The Journal of
  Chemical Physics}\ }\textbf {\bibinfo {volume} {131}},\ \bibinfo {eid}
  {054106} (\bibinfo {year} {2009})}\BibitemShut {NoStop}%
\bibitem [{\citenamefont {Shin}\ and\ \citenamefont {Metiu}(1995)}]{Metiu}%
  \BibitemOpen
  \bibfield  {author} {\bibinfo {author} {\bibfnamefont {S.}~\bibnamefont
  {Shin}}\ and\ \bibinfo {author} {\bibfnamefont {H.}~\bibnamefont {Metiu}},\
  }\href
  {http://scitation.aip.org/content/aip/journal/jcp/102/23/10.1063/1.468795}
  {\bibfield  {journal} {\bibinfo  {journal} {J. Chem. Phys.}\ }\textbf
  {\bibinfo {volume} {102}},\ \bibinfo {pages} {9285} (\bibinfo {year}
  {1995})}\BibitemShut {NoStop}%
\bibitem [{\citenamefont {Abedi}\ \emph {et~al.}(2012)\citenamefont {Abedi},
  \citenamefont {Maitra},\ and\ \citenamefont {Gross}}]{Ali_JCP}%
  \BibitemOpen
  \bibfield  {author} {\bibinfo {author} {\bibfnamefont {A.}~\bibnamefont
  {Abedi}}, \bibinfo {author} {\bibfnamefont {N.~T.}\ \bibnamefont {Maitra}}, \
  and\ \bibinfo {author} {\bibfnamefont {E.~K.~U.}\ \bibnamefont {Gross}},\
  }\href@noop {} {\bibfield  {journal} {\bibinfo  {journal} {The Journal of
  Chemical Physics}\ }\textbf {\bibinfo {volume} {137}},\ \bibinfo {eid} {22}
  (\bibinfo {year} {2012})}\BibitemShut {NoStop}%
\bibitem [{\citenamefont {Agostini}\ \emph {et~al.}(2015)\citenamefont
  {Agostini}, \citenamefont {Abedi}, \citenamefont {Suzuki}, \citenamefont
  {Min}, \citenamefont {Maitra},\ and\ \citenamefont {K.~U.}}]{AASMG2015}%
  \BibitemOpen
  \bibfield  {author} {\bibinfo {author} {\bibfnamefont {F.}~\bibnamefont
  {Agostini}}, \bibinfo {author} {\bibfnamefont {A.}~\bibnamefont {Abedi}},
  \bibinfo {author} {\bibfnamefont {Y.}~\bibnamefont {Suzuki}}, \bibinfo
  {author} {\bibfnamefont {S.~K.}\ \bibnamefont {Min}}, \bibinfo {author}
  {\bibfnamefont {N.~T.}\ \bibnamefont {Maitra}}, \ and\ \bibinfo {author}
  {\bibfnamefont {G.~E.}\ \bibnamefont {K.~U.}},\ }\href@noop {} {\bibfield
  {journal} {\bibinfo  {journal} {J. Chem. Phys.}\ }\textbf {\bibinfo {volume}
  {142}},\ \bibinfo {pages} {084303} (\bibinfo {year} {2015})}\BibitemShut
  {NoStop}%
\bibitem [{\citenamefont {Elliott}\ \emph {et~al.}(2012)\citenamefont
  {Elliott}, \citenamefont {Fuks}, \citenamefont {Rubio},\ and\ \citenamefont
  {Maitra}}]{elliott2012universal}%
  \BibitemOpen
  \bibfield  {author} {\bibinfo {author} {\bibfnamefont {P.}~\bibnamefont
  {Elliott}}, \bibinfo {author} {\bibfnamefont {J.~I.}\ \bibnamefont {Fuks}},
  \bibinfo {author} {\bibfnamefont {A.}~\bibnamefont {Rubio}}, \ and\ \bibinfo
  {author} {\bibfnamefont {N.~T.}\ \bibnamefont {Maitra}},\ }\href@noop {}
  {\bibfield  {journal} {\bibinfo  {journal} {Physical review letters}\
  }\textbf {\bibinfo {volume} {109}},\ \bibinfo {pages} {266404} (\bibinfo
  {year} {2012})}\BibitemShut {NoStop}%
\bibitem [{\citenamefont {Luo}\ \emph {et~al.}(2013)\citenamefont {Luo},
  \citenamefont {Elliott},\ and\ \citenamefont {Maitra}}]{luo2013absence}%
  \BibitemOpen
  \bibfield  {author} {\bibinfo {author} {\bibfnamefont {K.}~\bibnamefont
  {Luo}}, \bibinfo {author} {\bibfnamefont {P.}~\bibnamefont {Elliott}}, \ and\
  \bibinfo {author} {\bibfnamefont {N.~T.}\ \bibnamefont {Maitra}},\
  }\href@noop {} {\bibfield  {journal} {\bibinfo  {journal} {Physical Review
  A}\ }\textbf {\bibinfo {volume} {88}},\ \bibinfo {pages} {042508} (\bibinfo
  {year} {2013})}\BibitemShut {NoStop}%
\bibitem [{\citenamefont {Luo}\ \emph {et~al.}(2014)\citenamefont {Luo},
  \citenamefont {Fuks}, \citenamefont {Sandoval}, \citenamefont {Elliott},\
  and\ \citenamefont {Maitra}}]{luo2014kinetic}%
  \BibitemOpen
  \bibfield  {author} {\bibinfo {author} {\bibfnamefont {K.}~\bibnamefont
  {Luo}}, \bibinfo {author} {\bibfnamefont {J.~I.}\ \bibnamefont {Fuks}},
  \bibinfo {author} {\bibfnamefont {E.~D.}\ \bibnamefont {Sandoval}}, \bibinfo
  {author} {\bibfnamefont {P.}~\bibnamefont {Elliott}}, \ and\ \bibinfo
  {author} {\bibfnamefont {N.~T.}\ \bibnamefont {Maitra}},\ }\href@noop {}
  {\bibfield  {journal} {\bibinfo  {journal} {The Journal of chemical physics}\
  }\textbf {\bibinfo {volume} {140}},\ \bibinfo {pages} {18A515} (\bibinfo
  {year} {2014})}\BibitemShut {NoStop}%
\end{thebibliography}%
 
\end{document}